\documentclass[times,twocolumn]{article}
\usepackage{amsmath,amssymb}
\usepackage{graphicx,rotating}
\usepackage{breqn}
\usepackage[colorlinks,bookmarksopen,bookmarksnumbered,citecolor=red,urlcolor=red]{hyperref}

\def\Jmin{J_{\mathrm{min}}}
\def\Jmax{J_{\mathrm{max}}}
\def\eps{\varepsilon}
\def\u{\boldsymbol{u}}
\def\Ro{\mathrm{Ro}}

\begin{document}


\title{A conservative adaptive wavelet method \\ for the shallow water equations on the sphere}

\author{M. Aechtner,  N.K.-R. Kevlahan and T. Dubos}



\maketitle

\begin{abstract}
We introduce an innovative wavelet-based approach to dynamically adjust the local grid resolution to maintain a uniform specified error tolerance.  Extending the work of \cite{DubKev13}, a wavelet multi-scale approximation is used to make dynamically adaptive the TRiSK model \cite{Ringler10} for the rotating shallow water equations on the sphere. This paper focuses on the challenges encountered when extending the adaptive wavelet method to the sphere and  ensuring an efficient parallel implementation using {\tt mpi}. The wavelet method is implemented in {\tt fortran95} with an emphasis on computational efficiency and scales well up to $O(10^2)$ processors for load-unbalanced scenarios and up to at least $O(10^3)$ processors for load-balanced scenarios. The method is verified using standard smooth test cases~\cite{Williamson} and a nonlinear test case proposed by~\cite{Galewsky04}. The dynamical grid adaption provides compression ratios of up to 50 times in a challenging homogenous turbulence test case.  The adaptive code is about three times slower per active grid point than the equivalent non-adaptive TRiSK code and about four times slower per active grid point than an equivalent spectral code. This computationally efficient adaptive dynamical core could serve as the foundation on which to build a complete climate or weather model.
\end{abstract}


\section{Introduction}
\subsection{Adaptive spherical shallow water models}
Geophysical flows are characterized by a wide range of time and space scales. Eddies, jets, currents and wave-packets are typical features that appear locally and involve small scales. It is also necessary for global circulation models to resolve large-scale features with length scales of thousands of kilometres. Because the location of the smallest dynamically active scales changes incessantly, an optimally efficient computational model should have a dynamically adaptive grid that tracks small scale features and ensure that numerical errors remain below a target value.  In other words,  the model should automatically adapt its resolution locally where required in order to resolve emerging small-scale features, or coarsen as these features dissipate.  Fixed nested and stretched grids \cite{Krinner97} have been used in weather forecasting and regional climate modelling. However the non-uniform grid resolution of these statically adaptive models is based on {\em a priori\/} knowledge of the solution which is not possible for strongly nonlinear and non-stationary flows. 

Dynamical adaptivity, where the grid is adapted automatically based on the solution and changes in time, is still a research topic in geophysical science, and has not yet been incorporated into operational general circulation models. The book by~\cite{Behrens06} gives an introduction and overview to adaptive modelling in atmospheric science. Pioneered by~\cite{Ska89} for weather models, dynamical adaptivity has been introduced for rotating shallow water models by~\cite{Jablonowski2009Blockstructured}  (block structured, finite volume, latitude-longitude). Several models (interpolation-based, spectral element, cubed sphere) were compared by \cite{StCyr2008Comparison}. Solutions of the shallow-water equations on statically \cite{Ringler2011Exploring} or dynamically  \cite{Bauer2013Simulation} stretched unstructured meshes have also been examined. More recently wavelets have been used for adaptive ocean modelling by~\cite{Reckinger11}, although this was a collocation method on the plane that does not conserve mass. The potential of dynamically adaptive numerical methods for global ocean and atmosphere modelling is still being explored.

\subsection{Contributions of this work and outline}
A conservative adaptive wavelet method for shallow water equations on a regular staggered hexagonal C-grid was recently introduced by~\cite{DubKev13}. This prototype method was implemented for regular planar geometry in \texttt{matlab} and demonstrated the potential of this dynamically adaptive method for simulating multiscale geophysical flows. The present work is a sequel to \cite{DubKev13} that extends the adaptive wavelet approach to the sphere and reimplements the algorithm in {\tt fortran95} and {\tt mpi} with the goal of achieving high computational efficiency and good parallel scaling.   

After introducing the general numerical model and algorithm in sections~\ref{sec:wavada}--\ref{sec:model}, section~\ref{sec:sphgeom} deals with the technical challenges and modifications of the algorithm due to the nonuniform discrete geometry on the sphere (e.g. the fact that triangular cells are no longer uniform or equilateral). The parallel implementation,  data-structure and strategies for optimizing computational efficiency are described in section~\ref{sec:impl}. This section also summarizes the strong and weak parallel scaling performance of the method. Sections~\ref{sec:veri} and \ref{sec:turb} verify the accuracy of the model for standard smooth test cases \cite{Williamson} and for a more complex nonlinear test case~\cite{Galewsky04}.  Finally, we consider the most challenging shallow water test case for dynamically adaptive methods: fully-developed homogeneous rotating turbulence on the sphere.

\section{\label{sec:wavada} Wavelets on the sphere}
\subsection{Wavelet spaces}
A function $f(x)$ defined on a domain $\Omega \subset \mathbb{R}^n$ may be approximated by a set of discrete basis functions $\phi_k^j(x)$,
\[ f(x) \approx \sum_{k \in \mathcal{K}(j)} f_k^j \phi_k^j(x), \]
where $j$ is the scale, $k$ is the position, $\mathcal{K}(j)$ is the index set of positions defining the basis functions at each scale $j$ and $f_k^j$ are the weights (called scaling coefficients). The larger the scale $j$ the finer and more accurate the approximation and the bigger the index set $\mathcal{K}(j)$. Alternatively, we can represent $f(x)$ in wavelet space in terms of the difference between successive levels of approximation $j$ and $j+1$, which is spanned by the set of wavelet functions $\psi_m^j (x)$,
\begin{dmath}
f_{\Jmax}(x) = \sum_{k \in \mathcal{K}(\Jmin)} f_k^\Jmin \phi_k^\Jmin(x) + \sum_{j=\Jmin}^{\Jmax-1} \sum_{m \in \mathcal{M}(j)}\tilde{f}_m^j \psi_m^j(x),
\end{dmath}
where $\mathcal{M}(j)$ is the index set of positions defining the wavelets at each scale $j$. Note that we require a coarse representation at scale $\Jmin$ and we have truncated the representation at a finest level of resolution $\Jmax$.  The basis functions $\phi_k^j(x)$ spanning each scale $j$ are called \emph{scaling functions\/} and the functions $\psi_m^j(x)$ spanning the difference space between representations at successive scales $j$ and $j+1$ (i.e. the interpolation error) are called \emph{wavelets\/}.

This wavelet multi-resolution analysis relies on the fact that the grids at two successive scales $j$ and $j+1$ are {\em nested\/}.  The index sets $\mathcal{K}(j)$ and $\mathcal{M}(j)$  refer to nodes on the grid, and hence each wavelet and scaling function $\psi_m^j(x)$ or $\phi_k^j(x)$ (and accordingly each coefficient $\tilde{f}_m^j$ or $f_k^j$) is uniquely associated with a particular node.  Due to the nesting property of the grids, the union of the index sets $\mathcal{K}(j)$ and $\mathcal{M}(j)$ equal the index set of nodes at the finer scale $j+1$,
\[ \mathcal{K}(j+1) = \mathcal{K}(j) + \mathcal{M}(j). \]
This relation reflects the fact that the wavelets span the difference in approximation spaces between successive scales $j$ and $j+1$. It is also the basis of adaptive wavelet methods since the wavelet coefficients $\tilde{f}_k^j$ measures directly the interpolation error associated with deleting a node $x_k^j$ from the grid.   Wavelet-based adaptivity is described in detail in section~\ref{sec:adapt}.

\begin{figure}
\centering
\includegraphics[width=0.7\columnwidth]{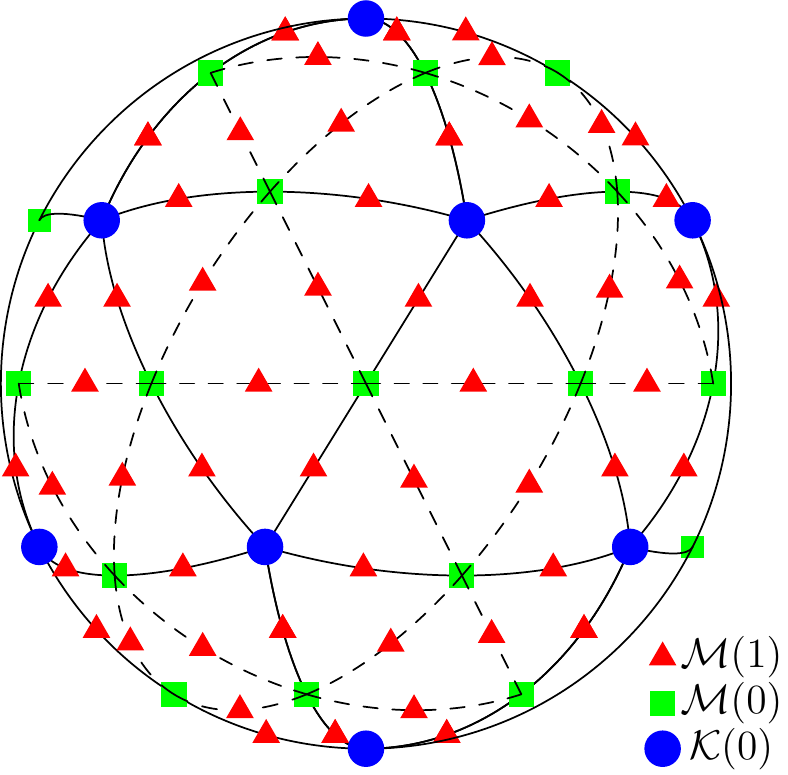}
\caption{\label{fig:sph_icosah} Nested grids on the sphere. An icosahedron projected to the sphere forms a coarse approximation (i.e. the blue nodes denoted by index set $\mathcal{K}(0)$).Refining this coarse grid via edge bisection and projection to the sphere produces a finer nested grid. $\mathcal{M}(0)$ are the new nodes added between level $j$ and $j+1$ and the finer grid has index set $\mathcal{K}(1) = \mathcal{K}(0) + \mathcal{M}(0)$.  Note that wavelets are located at nodes given by the index sets $\mathcal{M}(j)$ and scaling functions are located at nodes with index sets $\mathcal{K}(j)$. The red triangles are the new points added at the next finer level of approximation $j=2$.}
\end{figure}

Figure~\ref{fig:sph_icosah} shows three levels of nested grids $j=0,1,2$ on the sphere:  The round blue nodes are a coarse grid (level $0$, index set $\mathcal{K}(0)$). Together with the square green nodes $\mathcal{M}(0)$ they give the next finer level $j=1$ and satisfy the nested property $\mathcal{K}(0) + \mathcal{M}(0) =  \mathcal{K}(1)$.  Similarly, by adding  the red triangles we construct the next finer approximation level $j=2$ consisting of all nodes of any colour or shape.

\subsection{Discrete wavelet transform}
Second generation wavelets \cite{Swe96} allow the computation of the wavelet coefficients $\tilde{f}_m^j$ from the scaling coefficients $f_k^{j+1}$ by a discrete wavelet transform referred to as \emph{lifting}. Starting on the finest level $\Jmax$ and  working successively down to the coarsest level $\Jmin$, one computes 
\begin{equation}\label{eq:WTpred} 
\tilde{f}_m^j = f_m^{j+1} - \sum_{k \in \mathcal{K}_m^j} \tilde{s}_{km}^j f_k^{j+1} \quad \forall m \in \mathcal{M}(j).
\end{equation}
This is called the \emph{predict} step since the last term predicts (using interpolation) the values of the scaling coefficients $f_m^{j+1}$ that will be neglected at the coarser scale $j$. The $\tilde{f}_m^j$ are the wavelet coefficients and they measure the local interpolation error at scale $j$. Finally, one \emph{updates} the scaling function coefficients at scale $j$ by adding a linear combination of the neighbouring wavelet coefficients 
\begin{equation}\label{eq:WTupd} 
f^{j}_{k} = f^{j+1}_{k}+\sum_{m\in\mathcal{M}^{j}_{k}} s^{j}_{km} \tilde{f}^{j}_{m} \quad \forall k \in \mathcal{K}(j).
\end{equation}
This update step is used to improve properties of the transformation. In our case, we design the update step 
to ensure that the mean is conserved during refinement (i.e. prolongation) or coarsening (i.e. restriction) between different levels of resolution. Note that the predict and update weights $\tilde{s}_{km}^j$ and $s^{j}_{km}$ are zero except in a small neighbourhood of $l$ or $k$ respectively, i.e. they have finite and compact stencils.

Since, unlike first generation wavelets, second generation wavelets are constructed in physical space they can be designed for irregular domains and curved geometries.  Second generation wavelets were first developed for the sphere by~\cite{SS95}. The nested grid is generated by repeatedly bisecting the edges of an icosahedron, which forms the level $j=0$. (The blue dots in figure \ref{fig:sph_icosah} are the vertices of an icosahedron.)  Each bisection increases the scale $j$ by one. In practice, the coarsest level for the wavelet transform $\Jmin>0$ since the icosahedral grid is far too coarse to be practically useful.   As explained in section~\ref{sec:grid_optim}, the position of the grid points obtained must be adjusted slightly since after many such bisections the resulting triangular cells are increasingly non-uniform and far from the ideal case of equilateral triangles (at least near the 12 vertices and 30 edges of the original icosahedron).  We employ grid improvement techniques that globally optimize geometrical properties that are important for accuracy of the TRiSK finite volume/finite difference scheme we use to approximate the shallow water equations.

We now describe how filtering the wavelet coefficients can make this nested multiscale grid dynamically adaptive.

\subsection{\label{sec:adapt} Adaptivity using wavelets}
The discrete approximation of the function $f(x)$ at the finest scale $f_{\Jmax}$ can be compressed by removing (i.e. setting to zero) all those wavelet coefficients with modulus below a specified tolerance threshold $\eps$.  Due to the one-to-one correspondence between wavelet coefficients $\tilde{f}_m^j$ and grid points $x_m^j$ an adapted grid is obtained by including all grid points that correspond to active wavelet  coefficients. Since the wavelet coefficients are exactly the local interpolation error, this filtering ensures that error of the compressed function constructed by inverse wavelet transform on the adapted grid is at most $\eps$.

In addition to those significant wavelet coefficients above threshold, $|\tilde{f}_m^j| \geq \eps$, the adapted grid also includes all grid points on coarsest level $\Jmin$, and  all grid points that are adjacent in space (i.e. on the same level $j$) or in scale (on the next finer level $j+1$) to the significant wavelet coefficients. This allows dynamic adaptivity since adding nearest neighbours allows the grid to track energetic features as they move or develop smaller scales over one time step. This basic approach to wavelet adaptivity was first proposed by \cite{Lia90}.  For more details on wavelet-based adaptive numerical methods for partial differential equations we refer the reader to the review by \cite{Schneider/Vasilyev:2010}.

\section{\label{sec:model} Conservative wavelet method for the shallow water equations on the sphere}
\subsection{Discrete shallow water equations and flux restriction}
The evaluation of the free surface height perturbation $\delta h$ and horizontally averaged velocity $\u$ of a thin layer of fluid is described by the vector-invariant rotating shallow water equations
\begin{equation}
\frac{\partial \delta h}{\partial t} + \text{div} F = 0,
\label{eq:height}
\end{equation}
\begin{equation}
\frac{\partial \u}{\partial t} + q F^\perp + \text{grad} B = 0,
\label{eq:velo}
\end{equation}
with potential vorticity 
\[ q = \frac{\text{curl} \u + f}{h}, \]
thickness flux $F = h\u$, height $h = H + b + \delta h$, Bernoulli function $B = g \, \delta h + K$, kinetic energy $K = |\u|^2/2$,  Coriolis parameter $f$, bottom topography $b$, mean height $H$ and gravitational acceleration $g$.  $F^\perp$ is the flux perpendicular to the thickness flux $F$.

All differential operators are discretized using second-order finite volumes or finite differences as described in \cite{Ringler10} and the energy conserving variant is chosen for $qF^\perp$.  The prognostic variables $\delta h$ and $u$ are arranged in a staggered fashion:  the scalar values $h$ are located on nodes of the triangular grid and the vector field $\u$ is discretized by storing the normal velocity at the edge mid-points of the triangular cells located at the edge bisection. There are therefore three velocity components associated to each height variable and they are oriented parallel to each edge in a counter-clockwise fashion.  Thus, each triangular cell is associated with four discrete prognostic variables. Since we have two sets of variables on two different grids we require two distinct wavelet transforms: a scalar transform for height $h$ and a vector-valued wavelet transform for the velocity $\u$. These transforms are described in detail in \cite{DubKev13}. Note that, in contrast to \cite{DubKev13}, the prognostic variables $\delta h$ and $\u$ are represented as scaling coefficients (i.e. in physical space) instead of as wavelet coefficients (i.e. in wavelet space).

One time step consists of first computing $q$ and $K$ everywhere and $B$, $F$, and $qF^\perp$ on the locally finest level. Then the latter three quantities are restricted to coarser levels until they are available everywhere on the sphere. Finally, the gradient and divergence operators are evaluated and new $\delta h$ and $\u$ variables are computed from the trends.  At this point the time step is not yet completed. Partial wavelet transforms (for  height only) need to be computed to ensure consistency and mass conservation between levels, since prognostic variable are stored as scaling coefficients. The grid is then adapted based on the wavelet coefficients.  This means keeping, or if necessary adding, all grid points that either correspond to an active coefficients or are needed for the stencil of one of the differential operators, and removing the rest.  Since there are two types of wavelet coefficients (for $h$ and $\u$ respectively) the adaptation step also includes a consistency step that guarantees that active $h$ grid points (nodes) have active $\u$ points (edges) in their vicinity and {\em vice versa\/}.

The wavelet coefficient tolerance $\eps$ defined in section~\ref{sec:wavada} is not the actual threshold used for grid adaption. Instead it is a parameter that is set in order to control the error in the trend. In turn, $\eps$ determines the actual tolerances $\eps_h$ and $\eps_{\u}$ on the height and velocity wavelet coefficients. The relation between the thresholds for velocity wavelet coefficients, $\eps_{\u}$, and height wavelet coefficients, $\eps_h$,  to the trend tolerance $\eps$ depends on the regime and details can be found in \cite{DubKev13},
\begin{itemize}
\item Quasi-geostrophic regime: $\Ro = U/fL \ll 1$ 
\[ \eps_{\u} = f\, U\, L\, \mathrm{Ro}\, \eps^{3/2}, \]
\[ \eps_h = U\, \mathrm{Ro}\, \eps^{3/2}, \]

\item Inertia--gravity regime: $T \sim L/c$
\[ \eps_{\u} = c\, U\, \eps^{3/2}, \]
\[ \eps_h = U\, \eps^{3/2}, \]
\end{itemize}
where $U$ and $L$ are typical velocity and horizontal length scales, $c$ is the wave speed and Ro is the Rossby number.

For the discretization in time we use a four stage third-order strong stability preserving Runge--Kutta method that is stable up to CFL numbers of 2~\cite{Spiteri/Ruuth:2002}. The time step size is computed depending on the solution to guarantee stability 
\[
 \Delta t = \min \left(\frac{1}{\omega_{\text{max}}}, \left(\frac{\Delta x}{|\u|}\right)_{\mathrm{min}} \right)
\]

where $\omega_{\text{max}} = \left(\sqrt{f^2 + gh\pi/\Delta x}\right)_{\text{max}}$ is the maximum frequency supported on the grid. 

The maximum level $\Jmax$ may be determined implicitly by the tolerance $\eps$ or it can be set explicitly.  Allowing $\Jmax$ to be set by $\eps$ ensure spatially homogeneous error, but since an additional level is always added it also adds computational overhead.  It is also sometimes preferable to know the minimum grid resolution in advance of the simulation.  Similarly, the choice of the coarsest level $\Jmin$ also affects the efficiency of the method since retaining several completely filled levels results in unnecessary wavelet transform steps.(By a filled level we mean that the grid adaptation criteria force all grid points on a particular level to be retained.) Thus, efficiency requires that $\Jmin$ should not be less than the highest filled level.  Furthermore, if a particular level $j$ is almost entirely filled, it is still preferable to set $\Jmin \ge j$ since the extra nodes added are compensated by the gains of removing the lower level(s).  Although the choice of $\Jmin$ does not directly affect accuracy, increasing the minimum level can indirectly improve accuracy by improving grid quality (see \ref{sec:grid_optim}).  Typical values used in the test cases described below are $\Jmin=6$ (i.e. six dyadic refinements of the icosahedron) for a localized test function and  $\Jmin=7$ for a global test function if there are $O(10^6)$ d.o.f.~in the adaptive model. Table \ref{tab:sizes} shows different grid sizes and compares them to the equivalent spherical harmonic truncation limit $T$.
\begin{table}
\centering
\begin{tabular}{rrrrr}
\hline
$J$&   $N$   &     d.o.f & $\Delta x~[km]$ &   $T$\\
\hline
3 &       642&	   2,562 & 959.3	&    13\\
4 &      2,562& 	  10,242 & 479.6	&    25\\
5 &     10,242& 	  40,962 & 239.8	&    51\\
6 &     40,962&   163,842 & 119.9	&   101\\
7 &    163,842&   655,362 &  60.0	&   202\\
8 &    655,362&  2,621,442 &  30.0	&   404\\
9 &   2,621,442& 10,485,762 &  15.0	&   809\\
10&  10,485,762& 41,943,042 &   7.5	&  1619\\
11&  41,943,042&167,772,162 &   3.7	&  3238\\
12& 167,772,162&671,088,642 &   1.9 	&  6476\\
\hline
\end{tabular}
\caption{\label{tab:sizes} Number of bisection refinement levels  $J$  of the icosahedron, number of height nodes $N$,  total degrees of 
freedom (d.o.f.), average edge length $\Delta x$ and equivalent truncation limit for spherical harmonic spectral solvers $T$.  Level 
$J=12$ corresponds to a resolution of approximately one arc minute on the Earth.}
\end{table}

The model described above is inviscid, and the only source of dissipation is due to the wavelet adaptivity.  We also consider the shallow water equations with explicit dissipative terms added to both the height \eqref{eq:height} and velocity \eqref{eq:velo} equations,
\begin{eqnarray*}
\frac{\partial \delta h}{\partial t} + \text{div} F & = & \nu\,\text{div}\,\text{grad} (\delta h), \\
\frac{\partial \u}{\partial t} + q F^\perp & = & - \text{grad} \left(B + K\right) \\
& + & \nu \left(\text{grad}\,\text{div}\u - \text{grad}^{\perp}\,\text{curl} \u \right),
\end{eqnarray*}
where the new parameter $\nu$ is the viscosity.  The viscosity can be chosen to limit the minimum length scale, or to model dissipative mechanisms in the ocean or atmosphere (e.g. subgrid scale turbulent viscosity).

\subsection{\label{sec:grid_optim} Grid optimization}
Since the discretization of the differential operators from \cite{Ringler10} is second-order accurate  for equilateral triangles, but drops to first-order accurate when the triangles are far from equilateral, optimizing grid quality improves the accuracy of the solutions.  As explained earlier, this optimization is especially important for large numbers of scales (e.g. approximately for scales $J>6$) since the grid becomes increasingly distorted near the edges of the original icosahedron as the grid is successively refined by edge bisection.
\begin{figure}
\centering
\includegraphics[width=0.7\columnwidth]{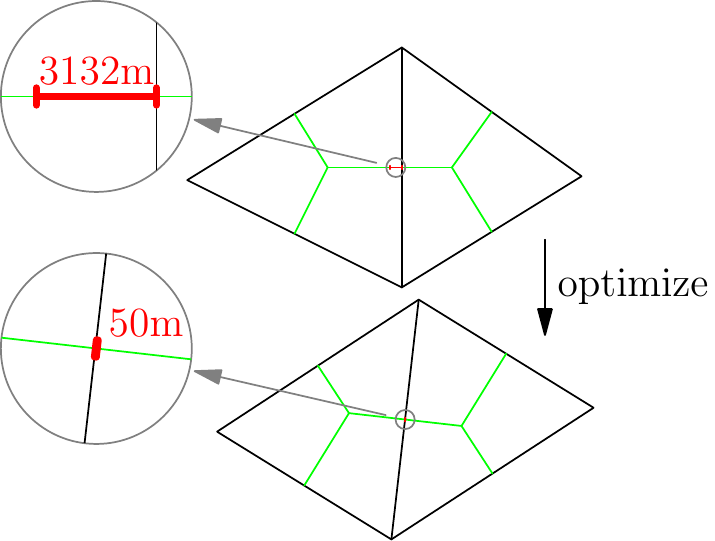}
\caption{Grid quality of simple bisection (top) and optimized grid (bottom). The offset (red, in meters at $J=7$) of edge bisection of primal (black) and dual 
grid (green) is reduced when the grid is optimized using the method of~\cite{Heikes13}.  The off-set error has been reduced by a factor of about 60. This results in a more accurate discretization of the differential operators in the TRiSK scheme.}
\label{fig:grid_optim}
\end{figure}

Figure \ref{fig:grid_optim} shows a section of the grid obtained by simple edge bisection (top) and the same section for an optimized grid (bottom) at $J=7$ obtained using the method of~\cite{Heikes13}. The approximation of the Laplacian operator is guaranteed to converge if the bisection of primal edge (black triangles) and dual edge (green hexagons) coincide. The distance (marked in red) between those two intersection points is an important measure for the grid quality.  On simple grids refined by edge bisection the Laplacian operator does not even achieve first-order convergence.

Optimized grids provided by~\cite{Heikes13} can be read into the model (this is currently the default method used in this paper). This approach seems to provide the best optimization and leads to a convergent Laplacian operator.  As an alternative, the grid optimization proposed by~\cite{Xu06} has also been implemented.  Advantages of this method is that it optimizes locally (rather than globally), is computationally inexpensive and easy to implement. However, while the grid quality is improved leading to lower error for a given resolution $J$, the Laplacian operator does not converge. In both cases the grid is first optimized on a coarsest level $\Jmin$ (determined by the physics of the problem and computational resources). Finer levels $j>\Jmin$ are obtained by edge bisection. This is necessary because the inter-scale restriction and prolongation operators used in the adaptive wavelet method require the grid points to be nested.

\section{\label{sec:sphgeom} New challenges from spherical geometry}
\subsection{General issues}
The main contributions of this work is to extend the planar model of~\cite{DubKev13} to allow for a non-uniform grid of non-equilateral triangles and to develop a highly efficient parallelized code and associated data structure.  In this section we consider the special challenges arising from the non-uniform discrete C-grid on the sphere.  In particular, due to fact that the weights and stencil geometry for discrete differential operators depend on position, the hexagonal grid on successive levels have complicated overlap regions, and the convergence behaviour of operators is affected. In the present method all areas are computed as spherical polygons, edges are spherical arcs and lengths are computed as arc-lengths on the sphere. In contrast to the plane, where only hexagonal cells occurred in the dual grid to the triangular primal grid, the sphere includes 12 exceptional pentagonal cells corresponding to the 12 vertices of the original icosahedron.

As in the planar version, the velocity requires a non-separable vector-valued wavelet transformation. This transformation involves interpolating the velocity at the mid-point of an edge of a fine level triangle $j+1$ from values of the edges on the coarser level $j$. Interpolation is carried out as linear combination of the velocity values on the coarse edges, where the (local) weights are pre-computed to guarantee second-order accuracy. At least six edges are required theoretically for second-order accuracy, but the model uses 12 edges in a symmetric stencil to gain stability and higher accuracy (see \cite{DubKev13} for more details). On the sphere every edge needs to compute and store its own weights, which are obtained by solving two $6\times 6$ systems of linear equations.

Combining \eqref{eq:WTpred} and \eqref{eq:WTupd} gives the action of the height restriction operator $R_h h_k^{j+1} = h_k^j$ as
\begin{dmath}
\label{eq:Rh}
h_k^j = h_k^{j+1} + \sum_{m \in \mathcal{M}_k^j} s_{km}^j h_m^{j+1} - \sum_{m \in \mathcal{M}_k^j} \sum_{k' \in \mathcal{K}_m^j} s_{km}^j \tilde{s}_{k'm}^j h_{k'}^{j+1}.
\end{dmath}
The scaling function coefficient at node $k$ on level $j$, $h_k^j$, corresponds to the average height on the hexagon whose centre is node $k$. The filter coefficients $\tilde{s}$ and $s$ are chosen such that the restriction from level $j+1$ to level $j$ conserves total height (i.e. conserves mass), 
\begin{dmath}
\sum_{k \in \mathcal{K}(j)} A_k^j h_k^j = \sum_{k \in \mathcal{K}(j+1)} A_k^{j+1} h_k^{j+1} \left(=\int_{Sphere} h\right).
\end{dmath}
Using \eqref{eq:WTupd} and \eqref{eq:WTpred} 
to express $h_k^{j+1}$ from $h_k^j$ and $\tilde{h}_m^j$, 
then setting to zero all but one coefficients among $h_k^j$ 
and $\tilde{h}_m^j$ yields the following conditions : 

\begin{eqnarray*}
A_{m}^{j+1} & = & \sum_{k\in\mathcal{K}_m^j} s_{km}^j A_k^j ,\\
A_{k}^{j} & = & A_{k}^{j+1}+\sum_{m\in\mathcal{M}_k^j}\tilde{s}_{km}^{j}A_{k}^{j+1}.
\end{eqnarray*}
These conditions are satisfied by letting 
\[\tilde{s}_{km}^j = \frac{A_{km}^{j+1}}{A_m^{j+1}}, \quad s_{km}^j = \frac{A_{km}^{j+1}}{A_k^j},\]
where $A_{km}^{j+1}$ is the area shared by the coarse level hexagon $A_k^j$ and the fine level hexagon $A_m^{j+1}$ (see figure~\ref{fig:Fcorr} ). Note that partial areas $A_{km}^{j+1}$ cover the fine and coarse scale hexagons,
ensuring that $s_{km}^j$ and $\tilde{s}_{km}^j$ are indeed weights. Thus,  it is necessary to compute the areas of intersection of spherical polygons. Hexagonal cells (and pentagons) are subdivided into six (or five) triangles using the central point (i.e. the barycentre). Since the types of triangle intersections that can appear during the $A_{m,k}^j$ computation are only a subset of all possible intersection cases, the intersection computation is optimized to account only for cases that can occur. The points of intersection of triangle edges are computed as spherical arc (great circle) intersections.

As in the planar case \cite{DubKev13}, in order to guarantee mass conservation the fluxes need to be restricted, and the restriction operators must satisfy the commutation relation
\begin{equation} 
\label{eq:divRcommute} 
R_h \circ \text{div}_{j+1} = \text{div}_j \circ R_F.
\end{equation}
On the sphere the construction of a flux restriction operator $R_F$ that guarantees this commutation property for a given height restriction operator $R_h$ 
poses additional difficulties due to location-dependent discrete geometry and due to the problem of overlapping hexagons at successive levels described above.

We use the strategy proposed by \cite{DubKev13} to split the height and flux restriction operators into a basic and correction part,
\begin{eqnarray}
R_F & = & R_{F0} + R'_F. \label{eq:R_f}\
\end{eqnarray}
The more complicated basic and corrective flux restrictions $R_{F0}$ and $R'_f$ needed in spherical geometry are described in the following two subsections.
\begin{figure}
  \centering
  \includegraphics[width=\columnwidth]{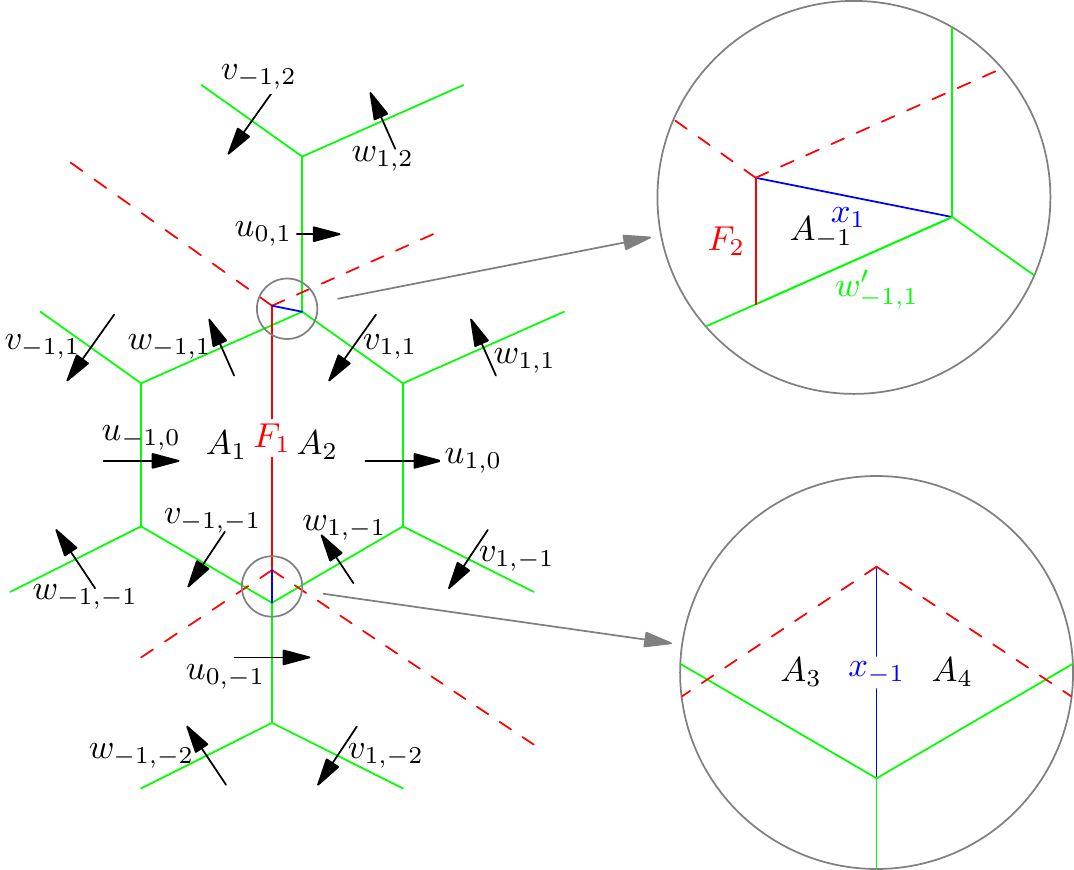}
  \caption{\label{fig:restr_F} Small overlap regions between hexagons at successive levels need to be accounted for when restricting the thickness flux.  The coarse level hexagons are red and the fine level hexagons are green. The inset figures show close-up views of the small overlapping areas due to the non-uniform C-grid structure on the sphere.}
\end{figure}

\subsection{Basic flux restriction} 
In the following, notation from figure~\ref{fig:restr_F} will be used, where all quantities (particularly $u, v, w, x$ and $F$) are integrated fluxes (total flux through an edge or part of an edge) except that $A$ stands for area.  As shown in figure~\ref{fig:restr_F}, the total area $A$ of the central green fine scale hexagon is decomposed as $A = A_1 + A_2 + A_3 + A_4$ according to the way it overlaps with the two adjacent red coarse scale hexagons sharing the solid red edge.

We assume that we are given all fluxes $u , v, w$ on the fine grid (green) and want to compute the flux through the solid red coarse edge $F = F_1 + F_2$ shown in figure~\ref{fig:restr_F}.  $F_2$ is the flux through the part of the coarse edge outside the green fine hexagon and $F_1$ is the flux through the part of the coarse edge inside the green fine hexagon. Here we consider the case where one end of the coarse edge is inside the fine (green) hexagon and the other end is outside.  In this way the procedure for both cases (ending inside and ending outside) is explained.  In the case that both, or no, edge is inside the fine hexagon the same procedure is simply applied at both ends.

The sum of fluxes entering the fine (green) hexagon on the left of the $F_1 + x_{-1}$ connection is defined as $F_{\text{in}}$ and the flux leaving on the right is defined as $F_{\text{out}}$:
\begin{eqnarray}
F_{\text{in}}   & = & -w_{-1,1} + u_{-1,0} -v_{-1,-1} + w_{-1,1}',\\
F_{\text{out}} & = & -v_{1,1} + u_{1,0} -w_{-1,1} + w_{-1,1}'.
\end{eqnarray}
The divergence theorem says that the average divergence of a vector field over an area $A$, $\text{div}_A$, is equal to the net flux through the boundary of $A$ divided by $A$,
\begin{equation} \label{eq:divflux}
\text{div}_A = \left( F_{\text{out}} - F_{\text{in}} \right) / A.
\end{equation}
Therefore, average divergence over the small area shown in figure~\ref{fig:restr_F} may be written as
\[
\text{div}_{A_1+A_2} = \frac{\left( F_1 + x_{-1} \right) - F_{\text{in}} }{A_1 + A_3} = \frac{F_{\text{out}} - \left( F_1 + x_{-1} \right)}{A_2 + A_4}.
\]
The above expression can be solved for the flux $F_1 + x_{-1}$, using $A_2 + A_4 + A_1 + A_3 = A$, to find
\[
F_1 + x_{-1}  = \frac{F_{\text{in}} \left( A_2 + A_4 \right) + F_{\text{out}} \left( A_1 + A_3 \right)}{A}.
\]
An expression similar to~\eqref{eq:divflux} also holds for the small triangle associated with area $A_{-1}$,
\[
 A_{-1}  \text{div}_{A_{-1}} = - F_2 - w_{-1,1}' - x_1.
\]
Solving for $F_2$ yields an expression for the flux $F_2$,
\[
 F_2 = - A_{-1} \text{div}_{A_{-1}}   - w_{-1,1}' - x_1.
\]
Combining these results, $w_{-1,1}'$ cancels, and we find that the total basic restricted flux $F_0$  corresponding to the action of the operator $R_{F0}$ on the fine scale fluxes is 
\begin{eqnarray}
F_0 &=& F_1 + F_2 \nonumber\\
 &=&\frac{A_2+A_4}{A} \left( -w_{-1,1} + u_{-1,0} -v_{-1,-1} \right) \nonumber\\
&& + \frac{A_1+A_3}{A} \left( -v_{1,1} + u_{1,0} -w_{-1,1} \right)  \nonumber \\
& & - A_{-1}\text{div}_{A_{-1}}   - x_1 - x_{-1}. \label{eq:restr_flux}
\end{eqnarray}
The remaining step is to compute the fluxes $x_1$ and $x_{-1}$ through the small boundaries shown in the zooms in figure~\ref{fig:restr_F}.  The flux through boundary $x_1$ is interpolated using the fluxes at fine edges on the upper half $u_{0,1}, v_{1,1}, w_{-1,1}, u_{-1,0}-v_{-1,1}, v_{-1,2}-w_{1,2}, w_{1,1}-u_{1,0}$. The flux on the lower half through $x_{-1}$ is found in the same way from the fluxes $u_{0,-1}, w_{-1,1},v_{-1,-1},u_{-1,0}-w_{-1,-1},w_{-1,-2}-v_{1,-2},v_{1,-1}-u_{1,0}$. We employ the interpolation formula used for interpolating velocities in~\cite{DubKev13}, which has the following advantages:
\begin{enumerate}
 \item Second-order accurate.
 \item Reliable in the case of equilateral triangles.
 \item Computationally efficient as it reuses components.
\end{enumerate}
(Note that the commutation relation \ref{eq:divRcommute} is satisfied irrespective of the interpolation formula used to compute the fluxes $x_1$ and $x_{-1}$.)  This completes the computation of the restricted flux obtained from the basic operator $R_{F0}$ in equation~(\ref{eq:R_f}).  We now explain how to compute the corrective part $R'_F$  of the flux restriction in equation~(\ref{eq:R_f})  in order to obtain the full flux restriction operator $R_F$. 

\subsection{Corrective flux restriction}
The operator $R'_F$ that guarantees the commutation property \eqref{eq:divRcommute} can be computed from the hexagon intersection areas above, where $F'$ is the part of the restricted flux obtained from the corrective operator $R'_F$. 

\begin{figure}
\includegraphics[width=\columnwidth]{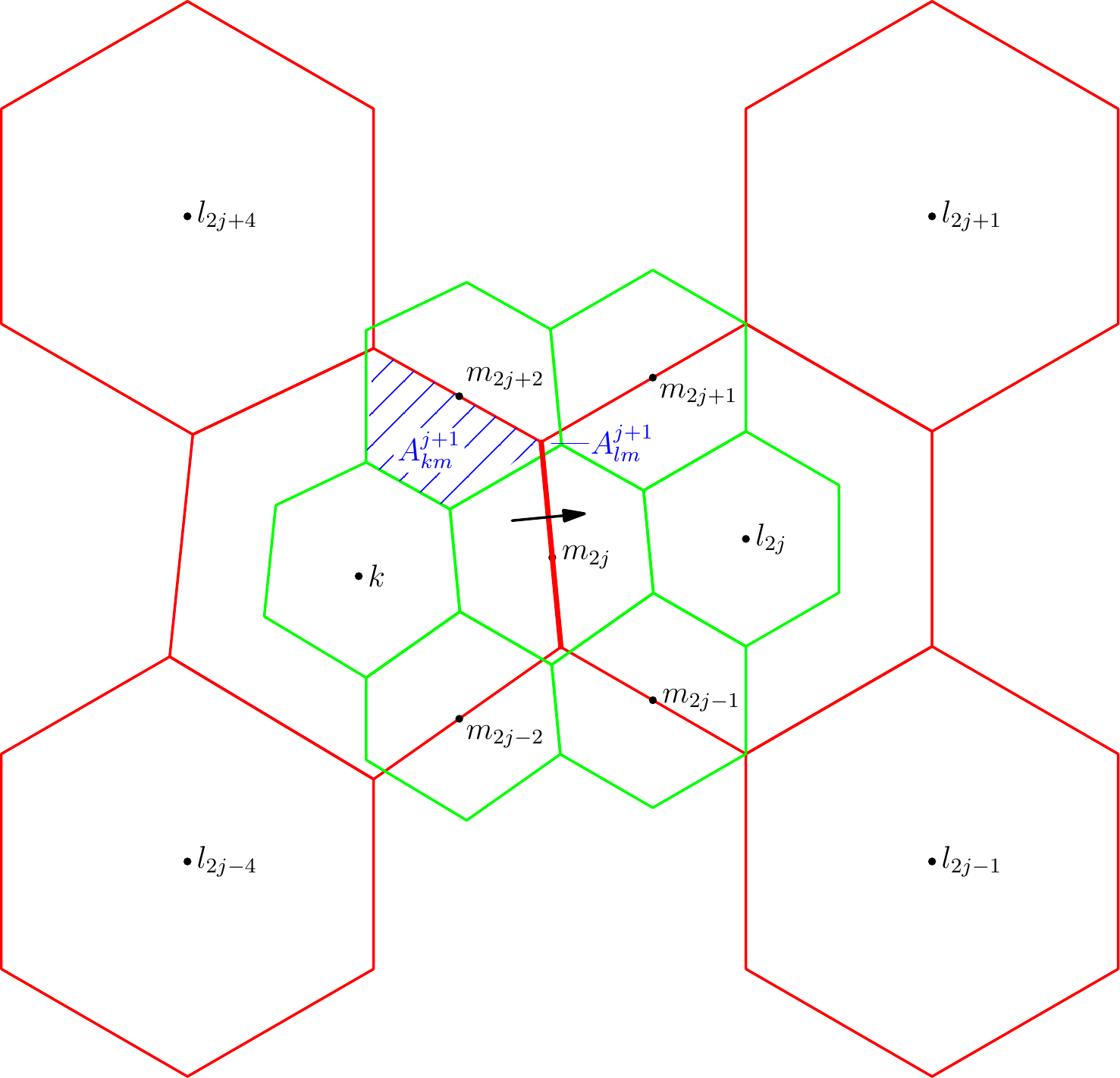}
\caption{\label{fig:Fcorr} Arrangement of fine and coarse scale height nodes used in the calculation of the corrective flux restriction through coarse edge $k l_{2j}$ (indicated by the arrow). The figure also shows the two partial areas $A_{km}^{j+1}$ and $A_{lm}^{j+1}$ used in the calculation of the flux restriction ($A_m^{j+1}$ is the complete fine scale green hexagon with partial areas $A_{km}^{j+1}$ and $A_{lm}^{j+1}$).}
\end{figure}

We assume that the hexagonal cell $k$ has $N$ edges (where $N=5$ for pentagons and $N=6$ for hexagons). The nearest
neighbour fine scale neighbours are denoted by $m_{0}, m_{2},\ldots,m_{2N-2}$ and the second
nearest neighbour fine scale neighbours are denoted by $m_{1}, m_{3},\ldots,m_{2N-1}$.  The
nearest neighbour coarse scale neighbours are denoted by $l_{0}, l_{2},\ldots l_{2N-2}$ and the
second nearest neighbour coarse scale neighbours are denoted by $l_{1}, l_{3},\ldots,l_{2N-1}$. They are arranged in such a way that:
\begin{itemize}
\item 
$m_{2j}$ is the midpoint of the edge joining nodes $k$ and $l_{2j}$ ; the second
nearest coarse neighbours of $m_{2j}$ are $l_{2j-2},\, l_{2j+2}$,
\item 
$m_{2j+1}$ is the midpoint of the edge joining nodes $l_{2j}$ and $l_{2j+2}$ ; the
second nearest coarse neighbours of $m_{2j+1}$ are $k$ and $l_{2j+1}$.
\end{itemize}
The arrangement of the nodes, points and edges used the calculation of the corrective flux restriction is shown in figure~\ref{fig:Fcorr}.

Using the notation in figure~\ref{fig:Fcorr}, the definition of the height restriction $R_h$ and the relation between the cell areas at the coarse and fine scales, the corrective part of the restricted flux $F'$ for the cell $k$ is given by
\begin{dmath}
\label{eq:ho_flux}
F' =  \sum_{j=0}^{N-1}  (km_{2j}l_{2j})+(km_{2j+2}l_{2j})+(km_{2j-2}l_{2j})+(km_{2j+1}l_{2j})+(km_{2j-1}l_{2j})
 +\frac{1}{2}(km_{2j+1}l_{2j+1}) + \frac{1}{2}(km_{2j-1}l_{2j-1}) + \frac{1}{2}(l_{2j+4}m_{2j+2}l_{2j}) + \frac{1}{2}(l_{2j-4}m_{2j-2}l_{2j})),
\end{dmath}
where
\begin{equation}
\label{eq:kml}
(kml) = \frac{A^{j+1}_{km} A^{j+1}_{lm}}{A^{j+1}_m} \left( c^{j+1}_k - c^{j+1}_l \right),
\end{equation}
and $c_k^{j+1} = \text{div}_k^{j+1} F_k^{j+1}$ is the divergence of the flux on the fine grid.  

In summary,  the restricted flux $R_F F_k^{j+1} = F_k^j$ is found by adding the basic restricted flux found using (\ref{eq:restr_flux})  to the corrective flux restriction found using  (\ref{eq:ho_flux},\ref{eq:kml}).  Note that to find the corrective flux restriction we must first calculate the local areas $A_{k}^{j}$ and $A_{km}^{j+1}$ associated with all active height nodes $x_k^{j+1}$ on the fine grid.  Using the height restriction~(\ref{eq:Rh}), it is relatively straightforward to verify that the complete flux restriction defined in (\ref{eq:restr_flux}, \ref{eq:ho_flux}) satisfies the commutation relation~(\ref{eq:divRcommute}).

\section{\label{sec:impl} Implementation and performance}
\subsection{General considerations}
The algorithm, which was previously implemented in {\tt matlab} for planar geometry by~\cite{DubKev13}, has been completely reimplemented in {\tt fortran95} with the goal of producing a code that is computationally efficient and scales well for parallel computation on large numbers of cpu cores. We have also made changes been made to the algorithm itself: the prognostic variables are stored in physical space instead of in wavelet space. Since most operators act in physical space this saves operations.

Because of the irregular geometry, most quantities (lengths, areas, weights, etc.) must be calculated individually for each computational element. Pre-computing these quantities increases memory use (which indirectly increases cpu-time), while computing them as needed considerably increases cpu-time.  We therefore decided which quantities to compute when a node becomes active and which to compute as needed in order to optimize total efficiency.  Additionally, quantities whose precision affects the mimetic properties (like mass conservation) are stored in double precision while values that are already lower accuracy due to truncation error, and which do not affect the mimetic properties, are stored in single precision.

\subsection{Hybrid data structure}
In terms of grid and data structures, the major difficulties arise from the spherical geometry (i.e. a non-regular domain) and the locally and dynamically adapted grid. In addition, the grid is staggered, rather than collocated.  Data can be associated either with triangles/circumcentres, with edges or with hexagons/nodes.  The  goal of this section is to construct a data structure that accommodates these properties and allows efficient computation of the most time-critical parts of the method: the differential and inter-scale operators.

A naive approach to deal with the triangular staggered grid on a non-regular domain would be to use a data structure where different grid entities are connected via coordinate references.  This has the disadvantage that finding second neighbours becomes expensive, additional data (for the references) has to be stored and communicated as the grid changes, and it is difficult to keep data locality under control.  A better solution is to use a hybrid data structure. 

\begin{figure}
\centering
\includegraphics[width=\columnwidth]{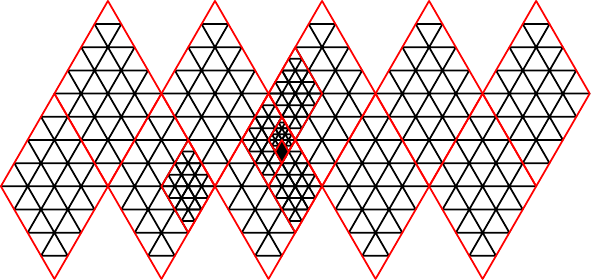}
\caption{\label{fig:hybrid_grid} Hybrid data-structure on an icosahedral grid. It is an irregular tree-like data structure with patches (red) as smallest element and a regular grid inside each patch. The figures illustrates an example where a small-scale structure in the centre caused adaptive grid refinement.}
\end{figure}

Ignoring the spherical geometry for the moment, the triangular staggered grid can be represented within a regular data-structure by grouping one node, two triangles and three edges into one computational element.  Then, unfolding the icosahedron, its grid is made up  of 20 triangles that can be grouped into 10 lozenges; (see figure \ref{fig:hybrid_grid}, disregarding the refined regions). Therefore, a grid resulting from refining an icosahedron can be divided into 10 sub-grids each of which can be stored and accessed in a regular fashion.  Note that at the edges of the lozenges the two adjacent regular grids of the original icosahedron are rotated with respect to each other. This is dealt with by surrounding the 10 lozenge sub-domains by ghost/halo cells, where values are not computed, but copied from their actual locations.  Alternatively, the nested levels of the adapted grid could be stored in a quad tree data-structure, but computational overhead during the neighbour search would be higher.  Neighbours could also be linked via references,  increasing the overhead in terms of memory and occasional cleaning and reference updating. 
\begin{figure}
\centering
\includegraphics[width=0.6\columnwidth]{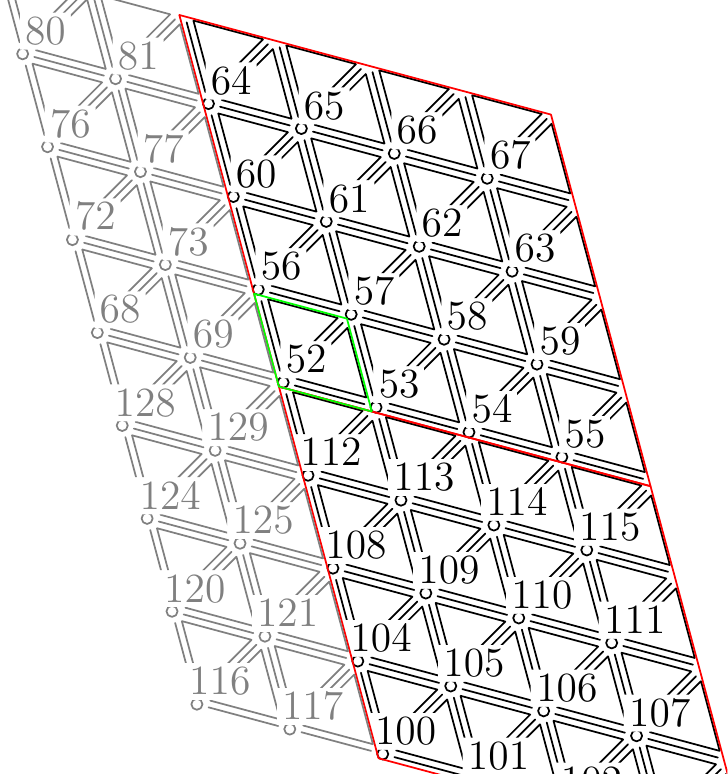}
\caption{\label{fig:zoom_grid} Section of the computational grid with ghost cells on the left.  A patch (red) is a regular grid of elements (green).  Each regular element is made up of of one node, two triangles and three edges}
\end{figure}

The best way to proceed in our case is to a hybrid data-structure: a combination of regular and irregular grids. The adapted data structure, the irregular part, uses patches as smallest elements. A patch constitutes a small regular grid. Inside a patch computations are efficient. A similar hybrid approach was used by  \cite{Behrens06} and \cite{HejRos10}.  In this way the references can be used to link patches to neighbours in space and scale, without introducing too much additional computational or memory overhead.  Since the granularity introduced by the patches involves computational overhead, a patch size that minimizes the total overhead needs to be found. Minimum patch sizes of $4 \times 4$ or $8 \times 8$ seem preferable, depending on the structure of the solution.  Figure~\ref{fig:hybrid_grid} shows an adapted grid with patch size $4 \times 4$ and $\Jmin = 0$.

Figure~\ref{fig:zoom_grid} shows a section of the grid with two $4 \times 4$ patches (red) which are located at the edge of a sub-
domain, and therefore have two rows of ghost cells to their left.  On the other sides there could be further patches (not shown) or there 
might not be any more patches on this level.  Taking element 52 as an example, neighbours inside the patch can be found from constant 
offsets (north: $+\text{patchsize}=+4$, east: $+1$). The other neighbours are located on neighbouring patches. Since every patch 
knows the patch indices of its eight neighbours, the neighbouring patches can be accessed to compute the element indices of the 
required elements. For example, the south offset is computed by finding the southern patch and then retrieving its element starting index 
(100). Then the offset is $100-52+\text{patchsize}(\text{patchsize}-1)$. Note that this offset stays the same for elements 52-55.
Note that the choice of the data structure does not affect the computed solution, only computational and memory efficiency. This means 
that on a patch only  active elements (as determined by the adaptive wavelet algorithm) are updated.

\subsection{Serial performance}
In this section we compare the performance of the serial version of our adaptive wavelet method with a similar non-adaptive single scale implementation of the TRiSK method~\cite{DYNAMICO} and a standard spectral implementation for the shallow water equations~\cite{Rivier:2002}.  All calculations were done on the same machine. 

Using the non-adaptive TRiSK implementation, a single time step takes $3.2 \times 10^{-7}\mathrm{s}$ per degree of freedom. The TRiSK simulation uses a uniform resolution corresponding to $\Jmax=8$ levels and $655\,362$ height nodes ($2.6\times 10^6$ total degrees of freedom) in table~\ref{tab:sizes}.  

We now compare the performance of the non-adaptive TRiSK code with a similar adaptive wavelet code.  The adaptive code has a maximum scale $\Jmax=10$ and uses 5 levels of refinement from $J=6$ to $J=10$.  (The $J=5$ grid is first optimized using the method of~\cite{Heikes13} before being used as the coarse level for the wavelet method.)  The total number of active height nodes in the $\Jmax=10$ adaptive wavelet method is 500\,962, roughly equivalent to the non-adaptive method.  This means the grid compression ratio is about 21 times for the adaptive wavelet method. The adaptive wavelet method  is 3.4 times slower per {\em active\/} node than the non-adaptive method.  Nevertheless, since the compression ratio is 21, the adaptive wavelet method is still about six times faster than the non-adaptive method in this case. Since the discretizations are identical, this result gives a good estimate of the total computational overhead due to the multiscale wavelet adaptivity.   Note that the overhead due to the wavelet adaptivity increases with the number of levels of refinement.  $j=5$ refinement levels corresponds to local refinement of 32 times, which is usually sufficient.

Spectral solvers are considered to be the most efficient non-adaptive solvers (at least for serial implementations), and so give a good lower bound on computational cost.  A time step with the spherical harmonics spectral solver \texttt{swbob}~\cite{Rivier:2002} takes  $2.2 \times 10^{-7} \mathrm{s}$ per degree of freedom for a truncation limit T341 with 465\,124 height nodes. Therefore, we can conclude that the serial adaptive wavelet TRiSK solver is about five times slower per active node than an equivalent spectral solver with a similar number of active height nodes.  However, when compression is taken into account, the adaptive wavelet method is about four times faster than the spectral method, but with a maximum resolution about four times finer.

It is to important to note that the cpu time per grid point is largely independent of the compression ratio (i.e. the proportion of active grid points). This is confirmed in figure \ref{fig:turb_compr_cpu} which shows that, while the compression ratio (on the right) varies by a factor of more than three, the cpu time (on the left) is approximately constant on average.  Thus, the computational overhead of the adaptivity should not depend sensitively on the degree of compression.

In conclusion, we find that if the compression ratio achieved is larger than about three then the adaptive model will be faster than an equivalent non-adaptive version.  As will be seen below (e.g. figure \ref{fig:turb_compr_cpu} right), even for statistically homogeneous flows like turbulence, typical compression ratios achieved are greater than 10--50.  Thus, in the serial case we expect that, in addition to achieving a uniform error and finer local resolution, the adaptive wavelet method should be three to fifteen times faster than the similar non-adaptive methods.  In special cases which are naturally very sparse, such as tsunami propagation, the adaptive code could be several hundred times faster than the non-adaptive code.

Parallelization is vital for high performance on large problems, and the following two sections explain the parallel algorithm and evaluate its strong and weak parallel scaling performance.

\subsection{Grid distribution and parallel algorithm}
Our goal is to run on at least several hundred cpu cores in parallel with a weak parallel scaling efficiency (see below) of at least 70--80\% in order to assess the potential of our code to run efficiently on an even larger numbers of cores, $O(10^3)$ to $O(10^4)$, in the future.  In particular, we need to identify where the parallel performance bottle-necks are.  

Starting from the ten lozenge sub-domains shown in figure~\ref{fig:hybrid_grid}, $10 \times 4^j$ sub-domains can be obtain by dyadic refinement $j$ times (i.e. using $j$ levels of adaptive resolution). The sub-domains are distributed in parallel over several cpu cores, where each core can have several domains.  Having several small domains, rather than one big domain, per core can improve cache efficiency through blocking. 

In an adaptive simulation each sub-domain will typically have a different number of active nodes, and thus requires a different amount of communication. The {\tt metis}~\cite{metis} graph partitioner is used to improve load balancing amongst the cores. {\tt Metis} allows us to assign weights to the graph nodes (representing the sub-domains) and graph edges (representing the number of connections between two neighbouring domains).  When the load distribution becomes uneven due to the dynamic adaptivity, the loads can be redistributed during check-pointing.

Every sub-domain is extended to hold as many ghost/halo cells as necessary for the various required operators. The values at the ghost 
cells are communicated as needed. Intra core communication is done by copying and inter core communication is done using {\tt mpi}. 
During grid adaption new patches are added as required and grid connectivity between domains is updated (via MPI as necessary). 
Communications occur at each trend computation and at each grid adaptation step, the latter being less frequent.  There is some 
leeway in the design of the communication pattern, which we use in order to do as much  communication as possible at each grid 
adaptation step so that the frequent communications are as light and fast as possible. In addition, critical  communications are carried 
out locally point-to-point rather than using global communication where possible. Where applicable communication is non-blocking so 
that the computations can continue while communication is taking place in the background.

\subsection{Parallel performance}
We  quantify parallel performance  with respect to both weak and strong scaling efficiency.  All calculations are performed on the {\sc sharcnet} cluster \texttt{requin}, which has  1541 AMD Opteron cores and a Quadrics Elan4 interconnect. Each processor has two cores and 8GB of local memory. 

Strong scaling efficiency $E_S$ is defined as 
\[ E_S = \frac{t_1}{N t_N}\le 1,\]
where $t_1$ is the time to do a given computation on one core, $N$ the number of cores  used and $t_N$ the time to do the computation $N$ cores. $E_S$ measures how cpu time decreases as the number of cores increased for a {\em fixed\/} problem size.  Ideally, $t_N$ should decrease proportionally with increasing $N$ since the processes can divide the (constant) work.  However, in practice when the portion of the total computation allocated to each core reaches a lower bound $t_N$ no longer decreases due to the non-parallelized part of the code (Amdahl's law) or because of communications overhead.  

\begin{figure}
\centering
\includegraphics[width=\columnwidth]{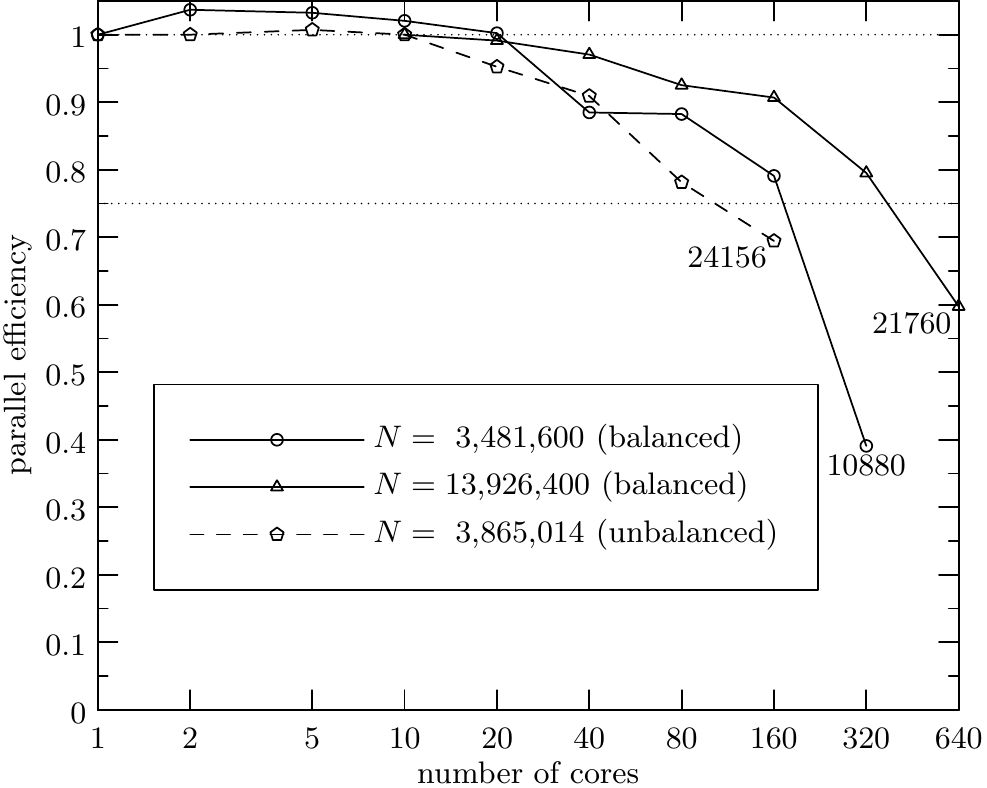}
\caption{\label{fig:strong_scal} Strong parallel efficiency scaling. Perfectly balanced (solid) and realistic turbulence test case (dashed). $N$ is the total number of degrees of freedom (four times the number of height nodes) and the numbers on the graph are active degrees of freedom per core for each case.}
\end{figure}

Figure~\ref{fig:strong_scal} shows the strong parallel scaling efficiency for a perfectly balanced load (solid line) and the turbulence test-case (dashed line). As expected, the unbalanced test case has a lower efficiency. The reason for the lower efficiency is explained below.  This result suggests that for both balanced and unbalanced problems strong parallel efficiency is acceptable for at least $10^2$ cpus.

\begin{figure} 
\centering
\includegraphics[width=\columnwidth]{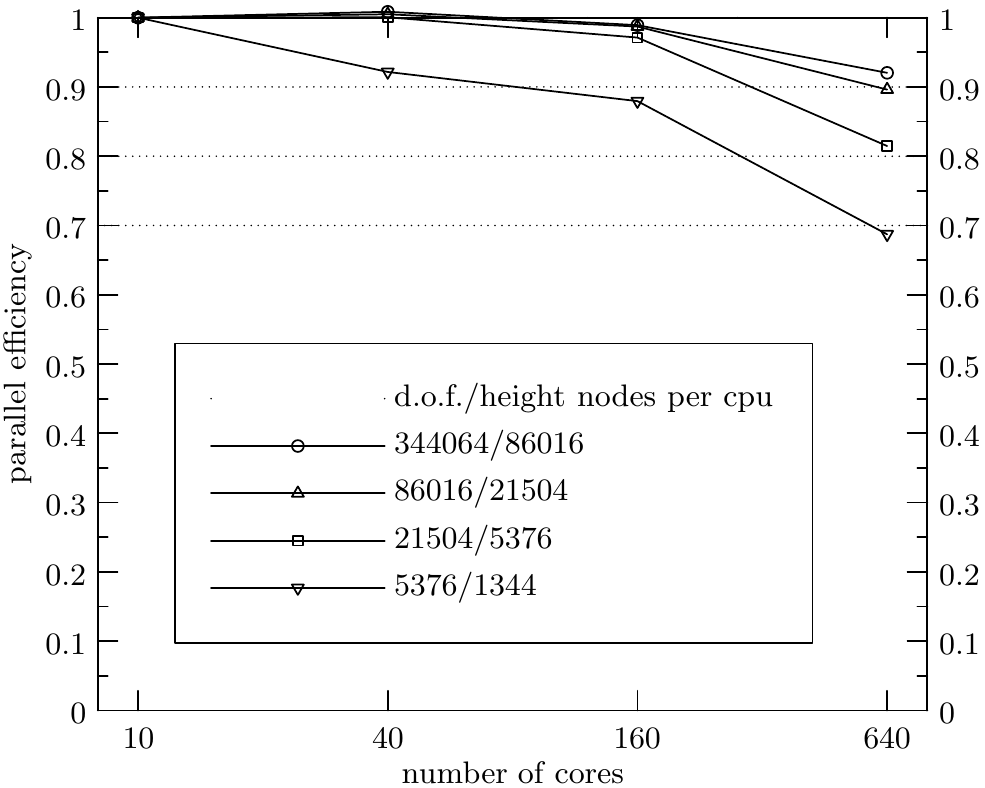}
\caption{\label{fig:weak_scal} Weak parallel efficiency scaling. Good performance is demonstrated for up to 640 cores (the maximum tested) with this perfectly balanced test case. Note that even with only 1344 height nodes per core this multilevel adaptive method achieves almost 70\% parallel efficiency on 640 cores.}
\end{figure}

In practice, weak scaling efficiency is a more useful measure since high performance codes are intended for large problems. To measure weak scaling efficiency the computation per core is kept approximately constant as number of processors is increased.  Weak scaling efficiency $E_W$ is defined as
\[ E_W = \frac{t_1}{t_N}\le 1, \]
where  $t_N$ is the time needed when running on $N$ processors. However,  unlike strong parallel efficiency, an efficient parallel code should maintain $E_W \approx 1$ independently of the number of cores used. Weak scaling is shown in figure \ref{fig:weak_scal} for a balanced test case. It demonstrates that good weak parallel efficiency can be achieved for at least 640 cores. In particular,  if at least there are at least 20\,000 height nodes per core the weak scaling efficiency is 90\% on 640 cores. Scaling for larger number of cpus could not be tested given resource limitations, although these results suggest that the code should have acceptable parallel performance for at least 1000 cores.

The adaptive algorithm requires a large number of communications, although only the inter-scale (interpolation and restriction) operators require communication with distant cores.  Most operators use the results of a previous operator available on neighbouring nodes.  For the TRiSK operators it is possible to communicate only the prognostic variables if we compute some intermediate quantities on ghost cells. The communications bottlenecks are the inter-scale operators: flux restriction, velocity interpolation and height interpolation.  After fluxes have been restricted from level $j+1$ to $j$, fluxes on level $j$ need to be communicated before restriction from $j$ to $j-1$ is possible (and similarly for the interpolation). This not only means that the number of communications grows with the number of levels, it also poses also a more difficult load balancing problem.  Now, in order to avoid processors waiting at the communication step for others to finish, the amount of work on level $j$ should be equally distributed amongst the cores for each level $j$. So not only is it desirable to have the same number of total elements on each core, but the elements should ideally be equally distributed at each individual level. This is a significantly more difficult goal to achieve, especially since the multiscale grid structure changes due to grid adaptation after each time step.

This communications bottleneck currently limits efficient {\em strong\/} parallel scaling to about $10^2$ cpus. There is, however, potential for improvement if multi-constraint load balancing is used and/or the parallelization is extended to a hybrid shared-distributed memory approach.

\section{Verification\label{sec:veri}}
In this section we verify the numerical accuracy, convergence and error control of the  adaptive wavelet method against several test cases.

We ran test cases 1, 2 and 6 from the standard shallow water test suite by~\cite{Williamson} for different thresholds and consequentially 
different number of active grid points in order to investigate convergence.  We also show results from the strongly nonlinear barotropic 
instability test case by~\cite{Galewsky04}.  All test cases use the following physical parameters appropriate for  the Earth: gravitational acceleration $g = 9.80616 \mathrm{\,ms^{-2}} $, radius $R = 6.37122 \times 10^{6} \mathrm{\,m}$ and rotation rate $
\Omega = 7.292 \times 10^{-5}\mathrm{\,s^{-1}}$.  Longitude $\lambda$ and latitude $\theta$ coordinates are related to Cartesian coordinates 
$(x,y,z)$ by
\[\theta = \arcsin (z/R), \quad \lambda = \text{atan2}(y/x). \]

\subsection{Advection: cosine bell (Williamson test case 1) and smooth bell}
This first test case considers only linear advection of a height field by a prescribed velocity. This case is a good  
test of the grid adaptation routines and grid stability.  The time-independent advecting velocity field is
\begin{eqnarray*}
u(\theta,\lambda) & = & U \left( \cos \theta \cos \alpha + \sin \theta \cos \lambda \sin \alpha \right),\\
v(\theta,\lambda) & = & -U \left( \sin \lambda \sin \alpha \right),
\end{eqnarray*}
with $U = 2 \pi R / (12 \text{ days})$. Two different initial conditions for the height perturbation are compared:
the cosine bell from test~case~1 in~\cite{Williamson} 
\[ h = \frac{H}{2}\left(1 + \cos(\pi r/L)\right),\]
and a smooth bell inspired by~\cite{Galewsky04}
\[ h = H e^{r^2/(r^2-2L^2)},  \]
with
\[r = R\, \text{arccos}\left(\cos \theta \cos \lambda \right), \]
and $H=1000$ and $L=R/3$. The second initial condition is included because the the cosine bell is only $C^1$ continuous at $r=L$. Because our grid adaptivity routine is based on second-order interpolation this non-smoothness at the edge of the cosine bell could potentially affect grid stability.  Both initial conditions constitute a localized bell that is advected once around the sphere. 
\begin{figure}
\begin{tabular}{cc}
\includegraphics[width=0.45\columnwidth]{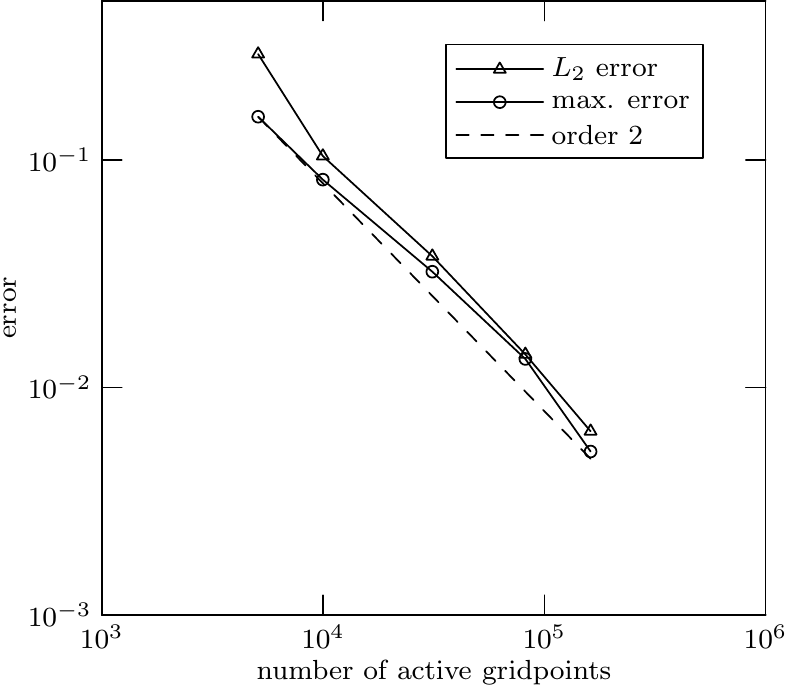} &
\includegraphics[width=0.45\columnwidth]{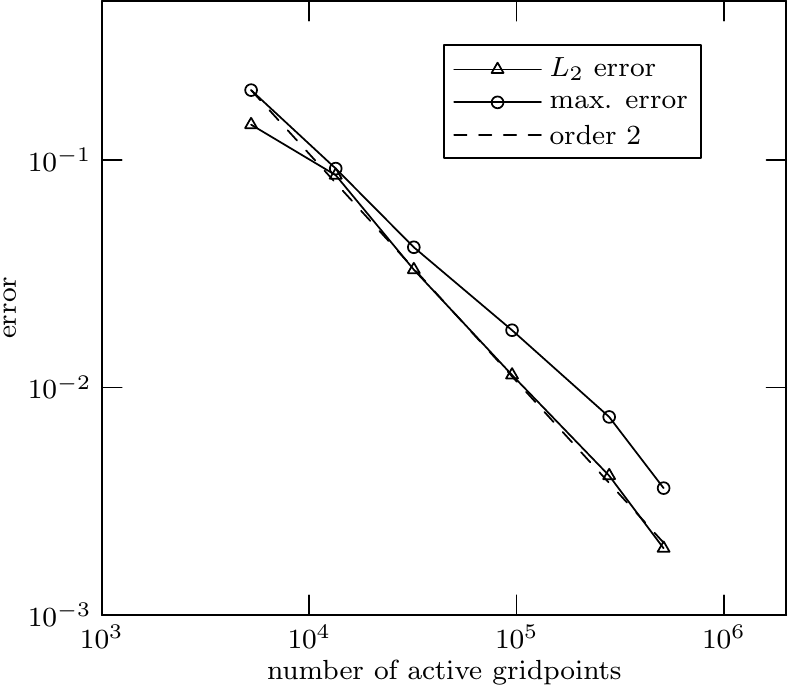}
\end{tabular}
\caption{\label{fig:will1conv} Errors with respect to analytic solution after 12 days (one rotation around the Earth) for the cosine bell (left), and smooth bell (right).}
\end{figure}
\begin{figure}
\centering
\includegraphics[width=0.7\columnwidth]{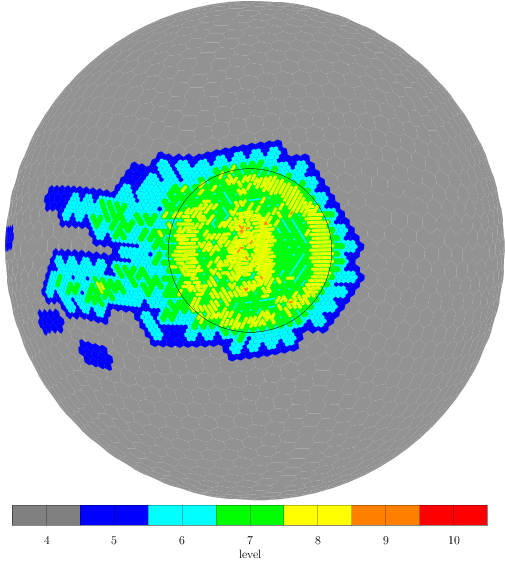}
\caption{\label{fig:will1grid} Grid after one rotation with cosine bell in the centre ($\epsilon=0.02, J_{\text min}=4$). The maximum level is determined by 
adaption routine.}
\end{figure}

Figure~\ref{fig:will1conv} (left) shows the convergence results for the cosine bell. The convergence of the error for increasing number of grid points corresponds to the expected second-order accuracy.  Recall that the number of active grid points is controlled by the tolerance $\epsilon$. The results for the smooth bell shown in figure~\ref{fig:will1conv} (right) are essentially the same as for the cosine bell.

The grid after one rotation (12 days) with the cosine bell, for a threshold for the trend of $\epsilon=0.02$, is shown in figure~
\ref{fig:will1grid}. The minimum level has been set to $\Jmin=4$. The maximum allowed level was set to $\Jmax=10$, but only levels up 
to $J=8$ are used. This shows that the actual maximum level is set by the tolerance $\epsilon$ (i.e. the simulation is fully adaptive in 
scale). The prescribed velocity in this figure goes from right to left. The grid is refined in the centre where the cosine bell is located and 
leaves a trace of refined grid that is gradually dissipates.  The smooth bell in fact shows a similar grid structure, and grid instability does 
not seem to be a problem for the non-smooth cosine bell.

\subsection{Test case 2: steady state geostrophic flow}
\begin{figure*}
\centering
\begin{tabular}{ccc}
\includegraphics[width=0.31\textwidth]{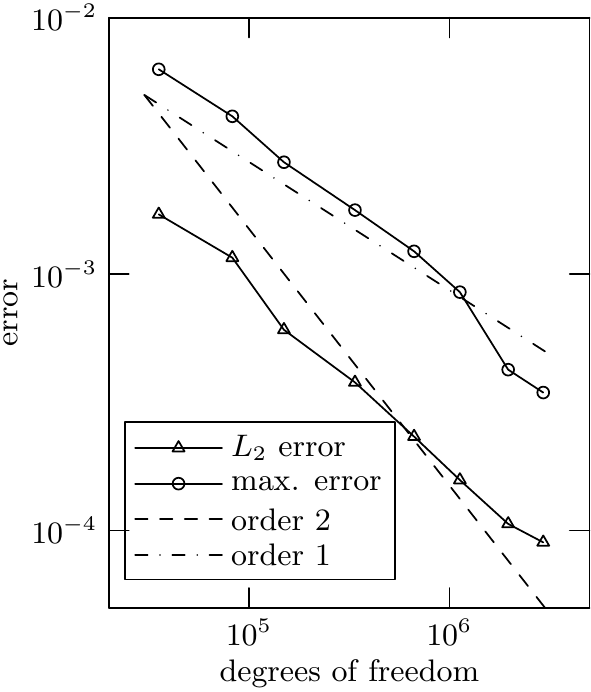} &
\includegraphics[width=0.31\textwidth]{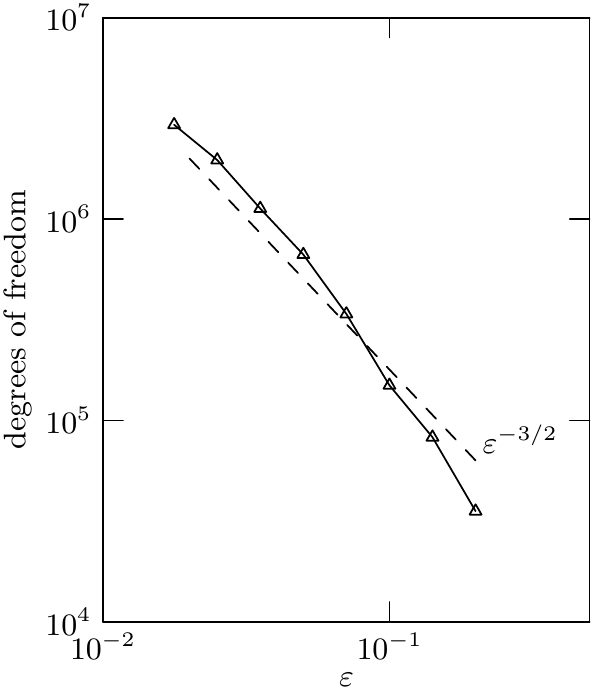} &
\includegraphics[width=0.31\textwidth]{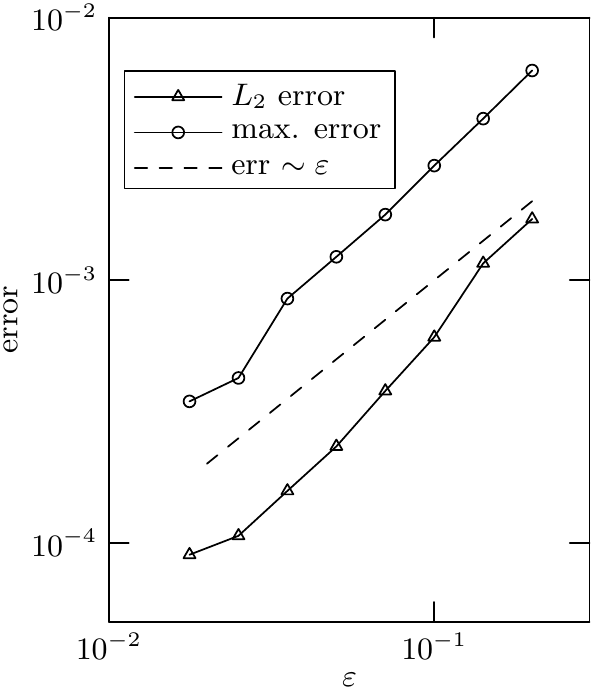} 
\end{tabular}
\caption{Test case 2 after 15 days. Errors for height (left), dependency of grid size on $\eps$ (middle) and error controlled by $\eps$ (right).}
\label{fig:will2conv}
\end{figure*}

The second test case uses the full shallow water equations. Height $h$ is defined by 
\[ gh = gH - \left(R\Omega U + \frac{U^2}{2}\right) \cos \theta,\]
and velocity as
\[ u = U cos \theta, \]
with $U = 2 \pi / ( 12 \text{ days})$ and $gH = 2.94 \times 10^4 \,\text{m}^2/\text{s}^2$.  The flow is in geostrophic balance so that the exact solution is equal to the initial condition at all times (steady solution). Figure \ref{fig:will2conv} (left) shows that the convergence of the global time integration error is approximately first-order accurate.  Figures~\ref{fig:will2conv} (middle and right) show, respectively, that the method is second-order accurate in space and that the accumulated error after 12 days is controlled by the tolerance $\epsilon$, as expected.

\subsection{Williamson test case 6: Rossby--Haurwitz wave}
Rossby--Haurwitz waves are a standard test case for the full shallow water equations.  The initial conditions are a non-divergent velocity field
\begin{eqnarray*} 
   u &=& a\omega \cos\theta + a K \cos^{R-1} \theta\, \left( R \sin^2 \theta - \cos^2 \theta\right) \cos R\lambda,\\
   v &=& -aKR \cos^{R-1} \theta\, \sin\theta\, \sin R\lambda,
\end{eqnarray*}
and a height chosen to ensure the flow is in geostrophic balance. This initial field rotates without change around the North--South axis.

\begin{figure}
\centering
\includegraphics[width=\columnwidth]{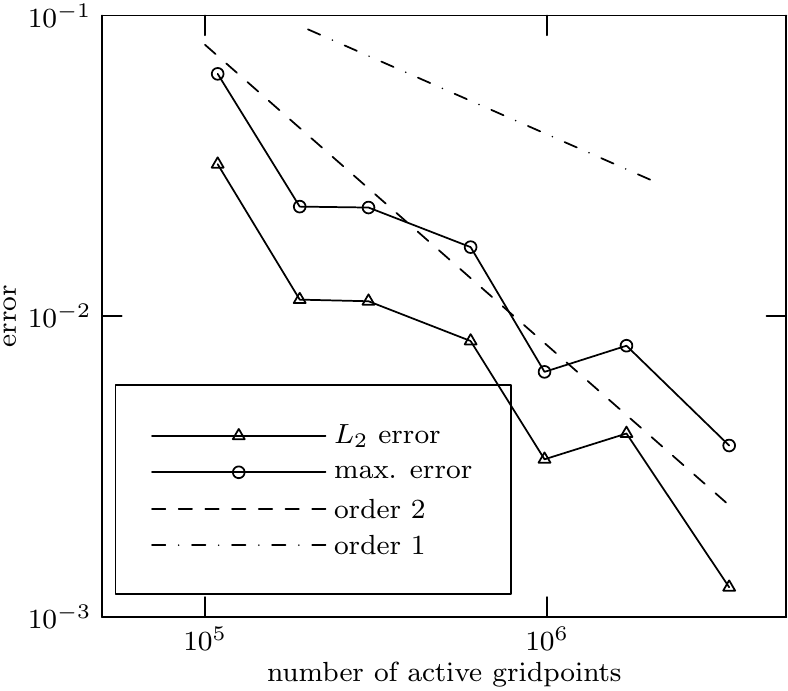}
\caption{\label{fig:will6conv} Test case 6. Error of the adaptive wavelet solution compared to the spectral solver \texttt{stswm} reference solution for Rossby--Haurwitz wave test case as a function of the number of active grid points.}
\end{figure}

Since analytical solutions are not available, solutions from the National Center for Atmospheric Research (NCAR) Spectral Transform Shallow Water Model \texttt{stswm} at resolution $T514$ are used as a reference.  Figure~\ref{fig:will6conv} shows that the convergence of the spatial error of the method is indeed approximately second-order for this full shallow water test case.

\subsection{\label{sec:galewsky} Galewsky disturbed jet}
\begin{figure}
\centering
\includegraphics[width=\columnwidth]{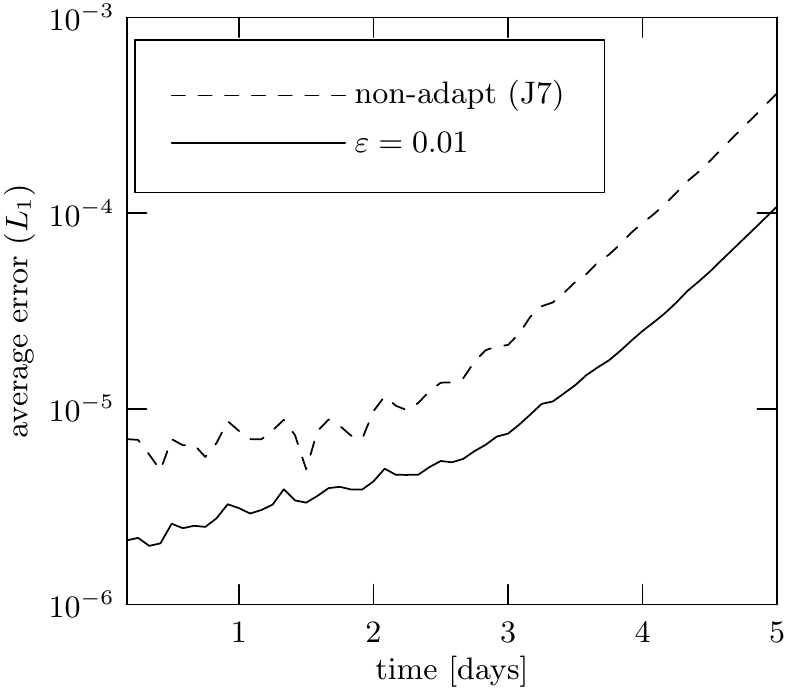}
\caption{\label{fig:gal_unpert} Unperturbed zonal jet test case~\cite{Galewsky04}. For similar numbers of active nodes the adaptive wavelet method  maintains consistently lower error than the non-adaptive TRiSK scheme.}
\end{figure}

The standard test cases above are supplemented by a strongly nonlinear test case proposed by~\cite{Galewsky04}: a zonal flow with a 
height disturbance that leads to an instability which eventually develops into turbulence.  As suggested in~\cite{Galewsky04},  the 
simulation is first run without the perturbation to assure that the numerical scheme is able to maintain balance for at least five days.  
Figure~\ref{fig:gal_unpert} shows the error in height for the first five days for the non-adaptive TRiSK scheme at resolution $\Jmax=7$ 
compared with results from the adaptive method with threshold $\epsilon$ chosen so that the total degrees of freedom are comparable 
to the non-adaptive simulation ($6\times10^5$).  These results show that, for a similar number of degrees of freedom and the same 
discretization scheme, the adaptive wavelet method maintains a significantly lower error (about three times lower). 

\begin{figure*}
\centering
\begin{tabular}{cc}
\includegraphics[width=0.48\textwidth]{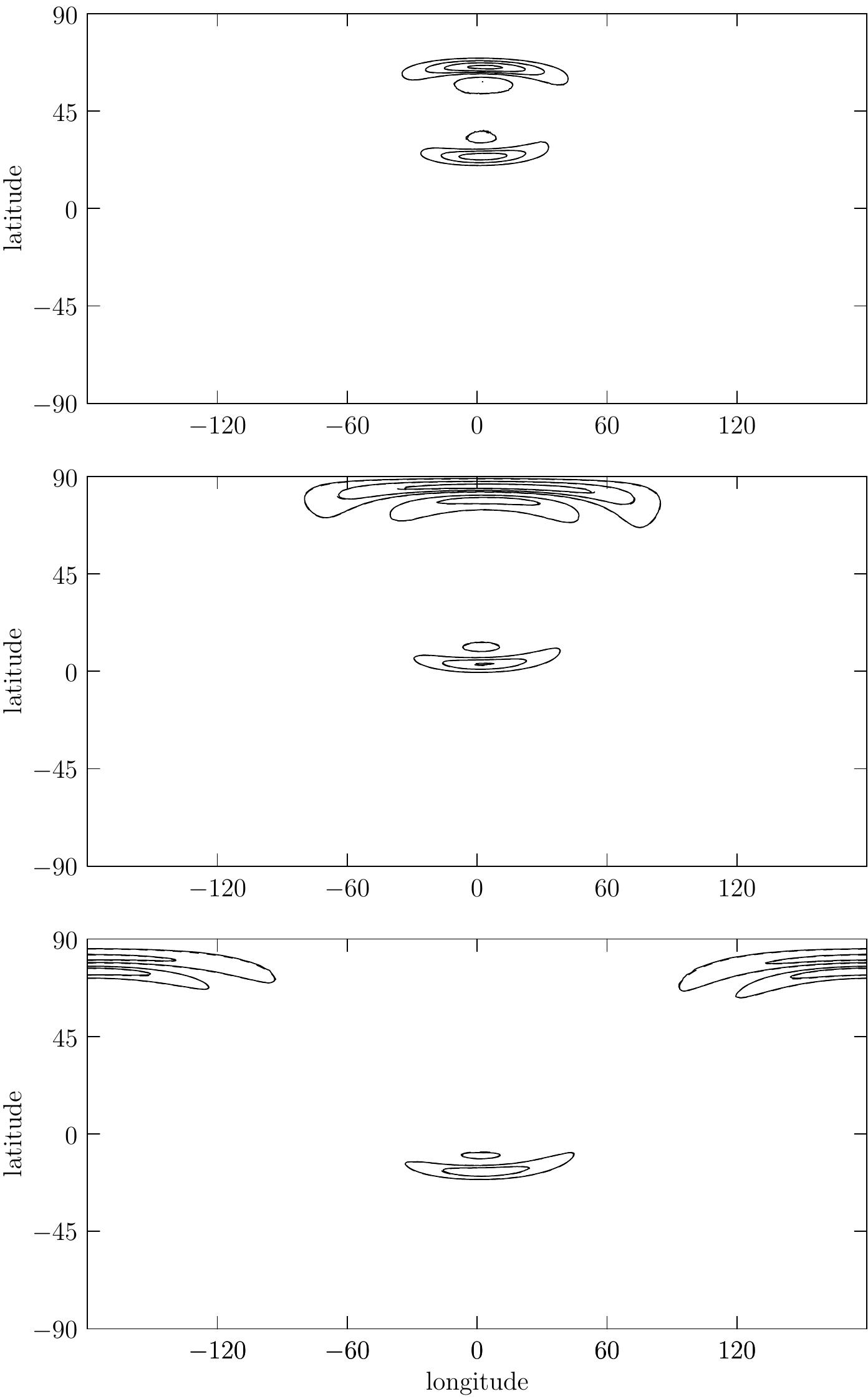}
\includegraphics[width=0.48\textwidth]{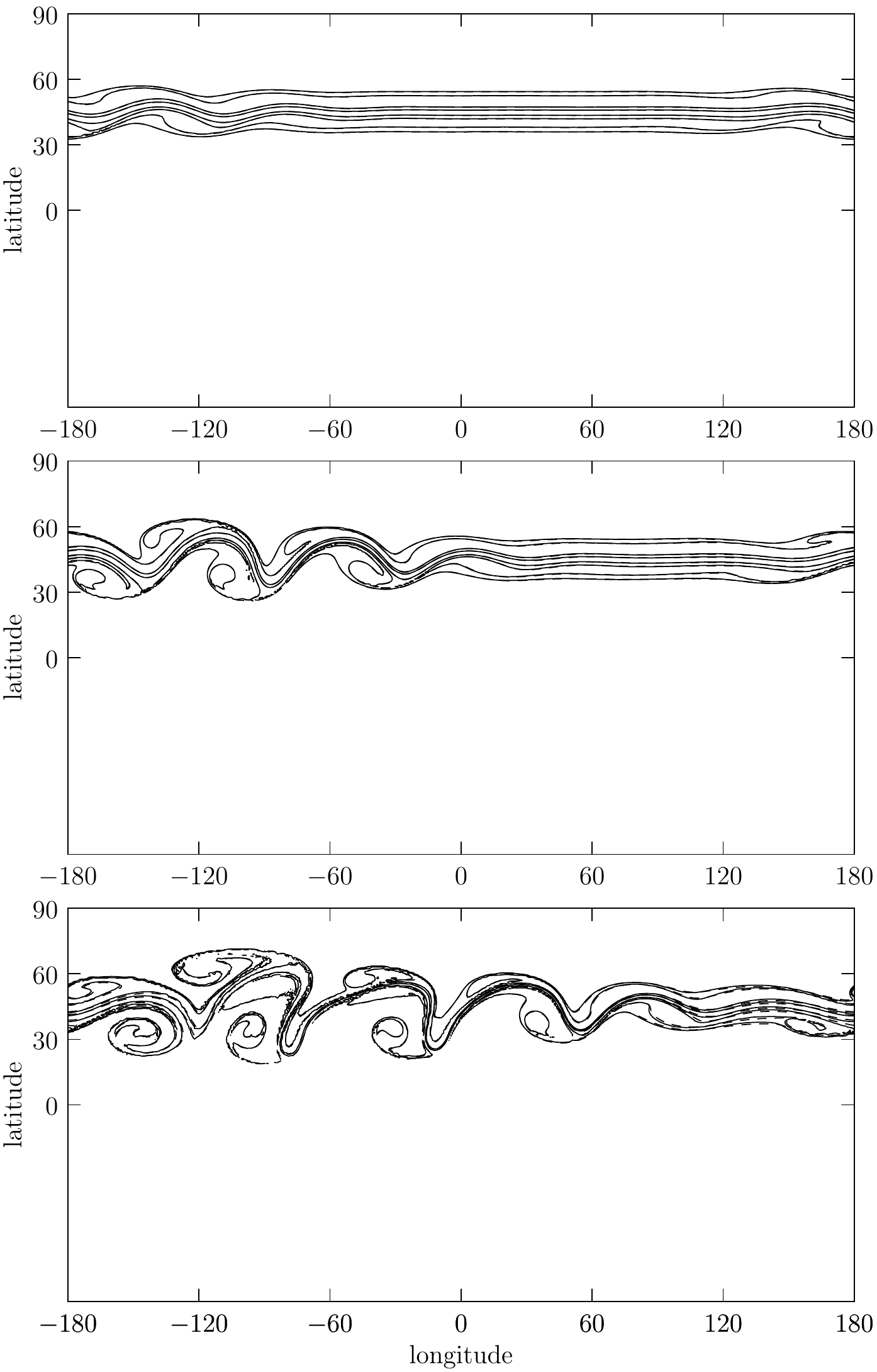}
\end{tabular}
\caption{\label{fig:galewsky_cont} \cite{Galewsky04} test case with tolerance $\eps = 5\times 10^{-3}$ and $\Jmax=9$.  Height perturbation at 2, 4 and 6 hours (left) and relative vorticity at 4, 5 and 6 days (right). The solution of $\Jmax=10$ non-adaptive reference simulation is dashed, but the lines are mostly indistinguishable.}
\end{figure*}

We now consider the results for the perturbed jet flow after the instability develops.  Results for tolerance $\eps=5\times 10^{-3}$, 
coarse scale $\Jmin=7$,  and finest scale $\Jmax=9$ are shown in figure~\ref{fig:galewsky_cont}. This simulation uses about 
$2\times10^6$ degrees of freedom, for a compression ratio of 5.25. The contours (solid) of the adaptive wavelet simulation nearly 
overlap with those of a reference simulation with the non-adaptive TRiSK scheme at the finer uniform resolution $\Jmax=10$ showing 
that the adaptive wavelet simulation is quite accurate, even for this highly nonlinear time-dependent test case.

We now consider one of the most challenging applications of a dynamically adaptive method: homogeneous isotropic rotating turbulence on the sphere.

\section{\label{sec:turb} Rotating shallow water turbulence on the sphere}
\subsection{Initial condition structure of solution}
As a final challenging test case closer to geophysically relevant applications, we consider initial conditions designed to generate shallow 
water turbulence.  The coarsest grid is at level $\Jmin=5$ and the finest level is determined by the tolerance $\epsilon$ (it turns out the 
finest level required is $\Jmax=10$). Both inviscid and viscous ($\nu = 10^4$) simulations are run with the same tolerance $\eps = 
5\times 10^{-2}$ corresponding to about $2\times 10^6$ degrees of freedom.  

\begin{figure}
\centering
\includegraphics[width=\columnwidth]{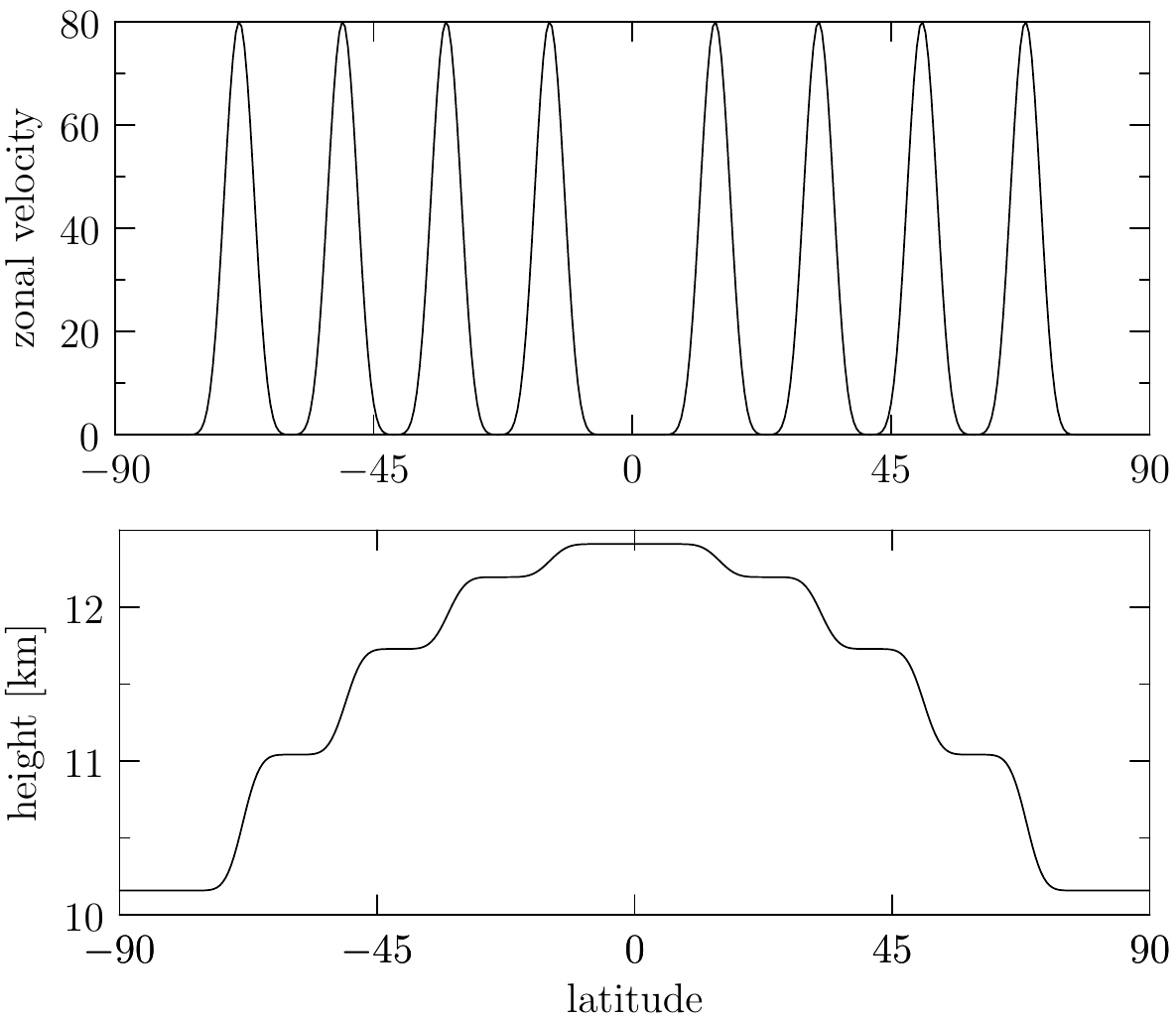}
\caption{\label{fig:turb_init_1d} Initial conditions for  zonal velocity (top) and height (bottom) for the turbulence test case.}
\end{figure}

\begin{figure*}
\centering
\includegraphics[width=\textwidth]{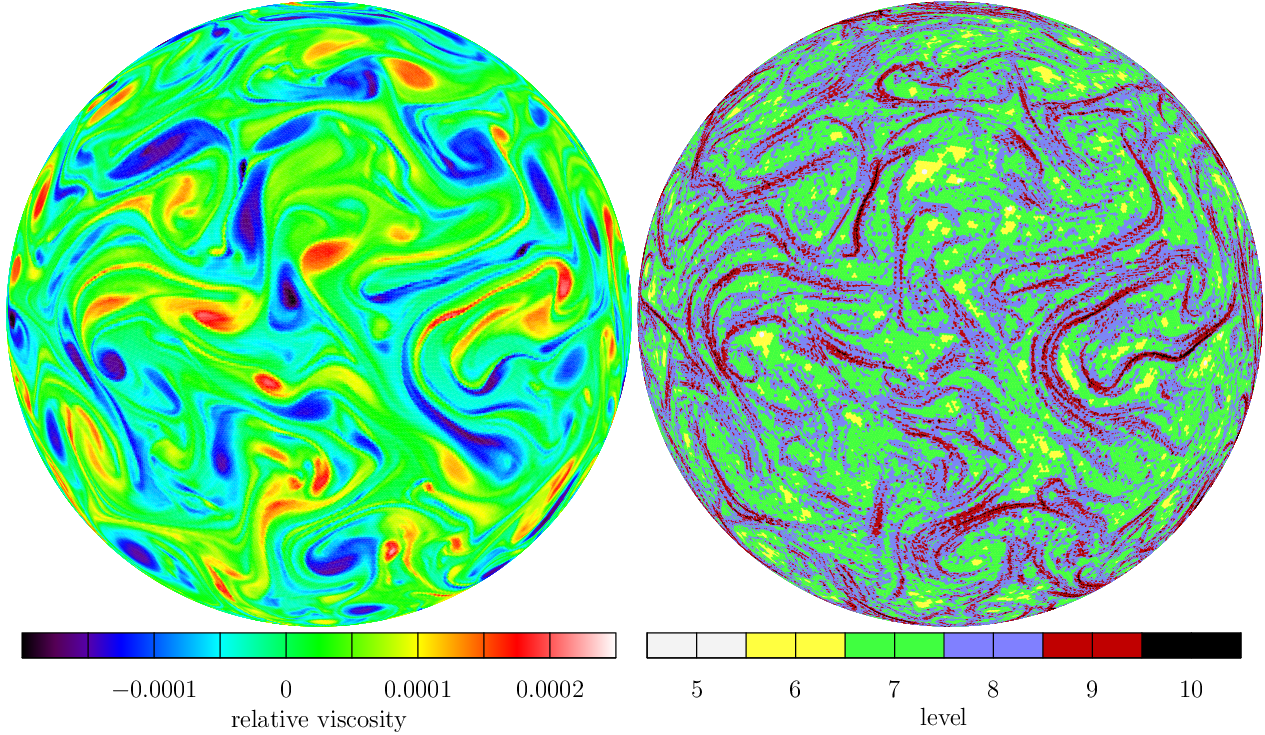}
\caption{\label{fig:turb_invi} Inviscid shallow water turbulence with tolerance $\eps=5\times 10^{-2}$ at time $t=132 \text{ h}$. Relative vorticity (left) and adapted grid (right). }
\end{figure*}

\begin{figure*}
\centering
\includegraphics[width=\textwidth]{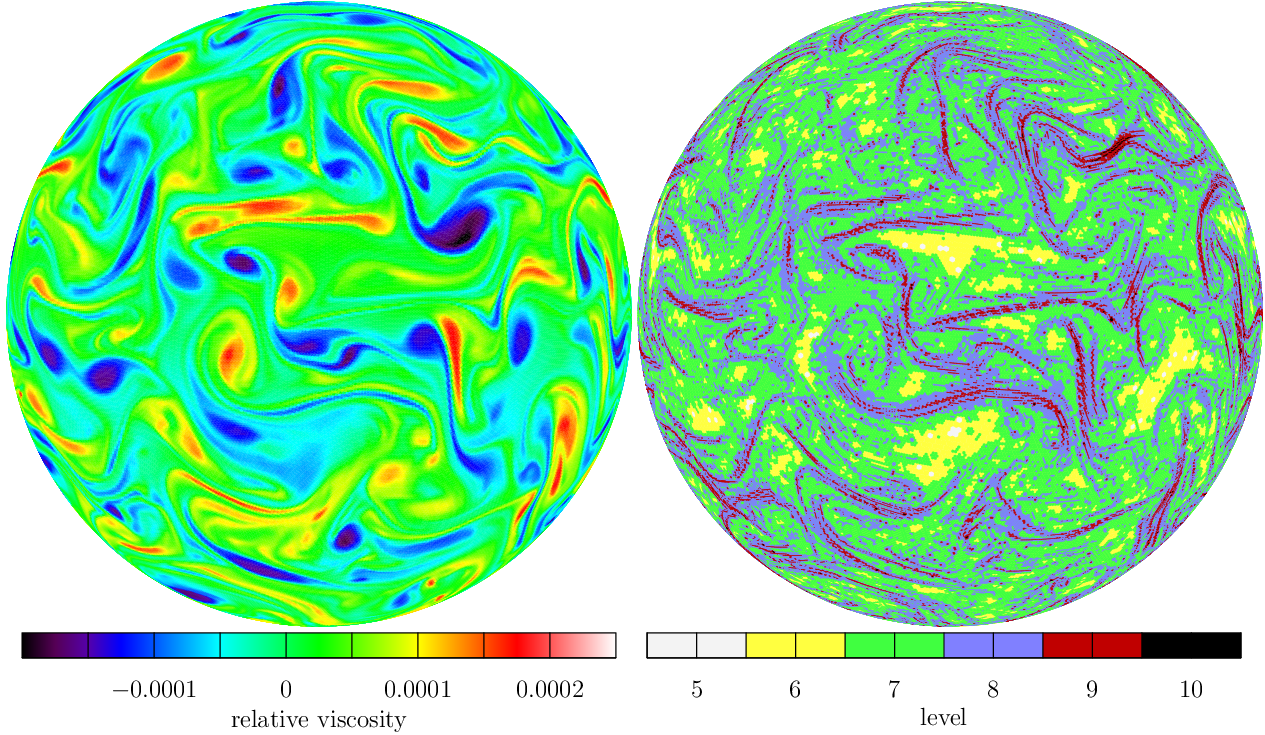}
\caption{\label{fig:turb_visc} Viscous shallow water turbulence with tolerance $\eps=0.05$, viscosity $\nu=10^4$ at time $t=132 \text{ h}$. Relative vorticity (left) and adapted grid (right).}
\end{figure*}

The initial condition is made up of several zonal jets similar to the zonal flow in section~\ref{sec:galewsky} arranged from North to South as shown in figure~\ref{fig:turb_init_1d}. Each zonal jet is perturbed to trigger an instability.   After two days vortices form on each of the jets that then interact to generate the approximately homogeneous and isotropic turbulence shown in figures~\ref{fig:turb_invi} and \ref{fig:turb_visc}.

Figures~\ref{fig:turb_invi} and \ref{fig:turb_visc} show the simulation results after 132 hours for the inviscid and viscous runs, 
respectively. The left hand figures showthe  relative vorticity and the right hand figures show the adapted grid. Each grid level is 
identified by a distinct colour. The most refined regions corresponding to the darkest colours, and are located near the intense vorticity 
filaments that characterize two-dimensional turbulence.

\begin{figure}
\begin{tabular}{cc}
\includegraphics[width=0.45\columnwidth]{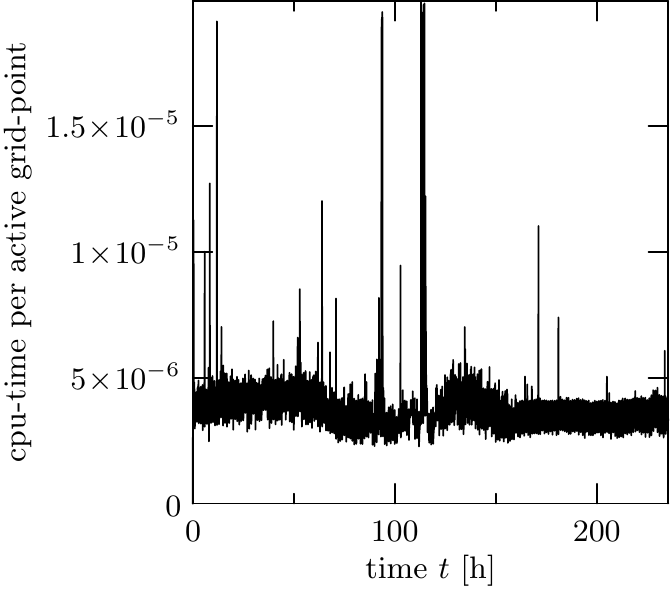}
\includegraphics[width=0.45\columnwidth]{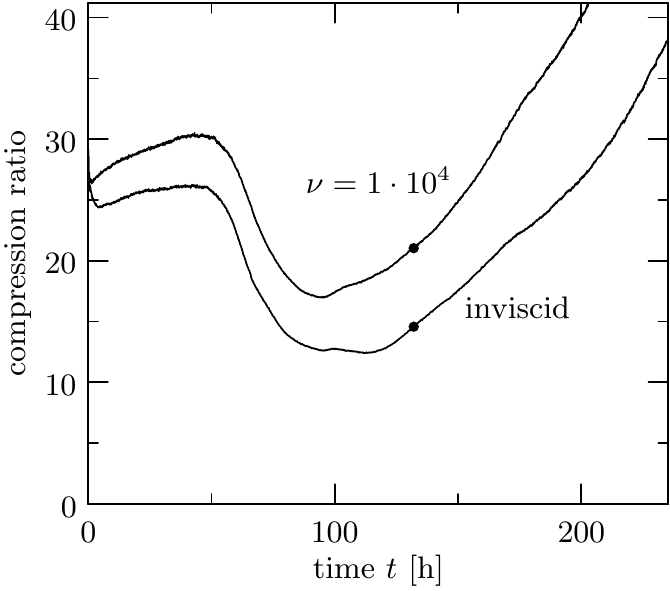}
\end{tabular}
\caption{\label{fig:turb_compr_cpu} Turbulence test case with tolerance $\epsilon=5\times 10^{-2}$). Cpu time per active grid point (left) and compression ratio based on the maximum scale $\Jmax=10$ (right).}
\end{figure}

Figure \ref{fig:turb_compr_cpu} shows that the compression ratio at $t=132$ hours is about 15 for the inviscid case (\ref{fig:turb_invi}) 
and 21 for the viscous case (\ref{fig:turb_visc}). It is important to note that at this time the compression ratio is at its lowest level since 
the turbulence is most intense (compared with both the initial conditions and dissipated flow at later times). Figure~
\ref{fig:turb_compr_cpu} also shows that the cpu time per active point remains roughly constant (left) even though the compression ratio 
changes significantly when turbulence first develops and then decays again (right).  This shows that there is no appreciable 
computational overhead associated with the degree of grid compression (sparse or dense grids).  Not surprisingly, the compression ratio 
is lowest when the flow is most turbulent. Nevertheless, the viscous adaptive wavelet code is still about four times faster than the 
spectral code and six times faster than the non-adaptive TRiSK code at this time for an equivalent maximum resolution. This result 
confirms that adaptive methods can still be advantageous for statistically homogeneous and isotropic flows, like fully-developed two-
dimensional turbulence.  
 
\subsection{Energy and spectrum}
\begin{figure}
\begin{tabular}{cc}
\includegraphics[width=0.45\columnwidth]{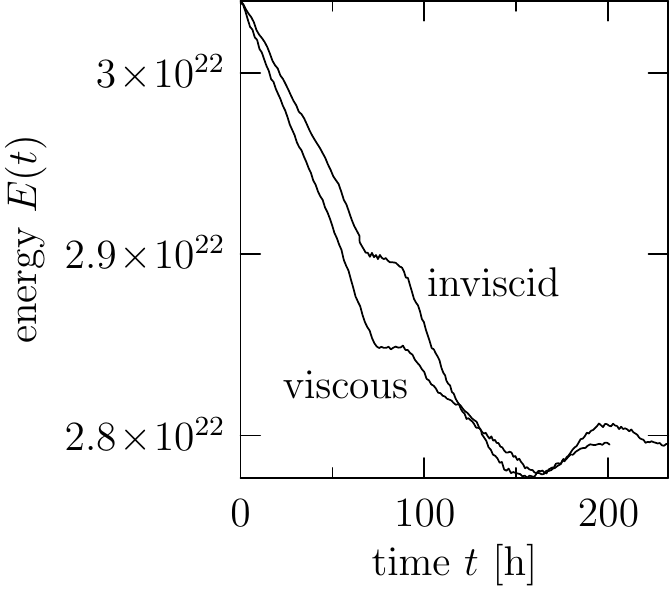}
\includegraphics[width=0.45\columnwidth]{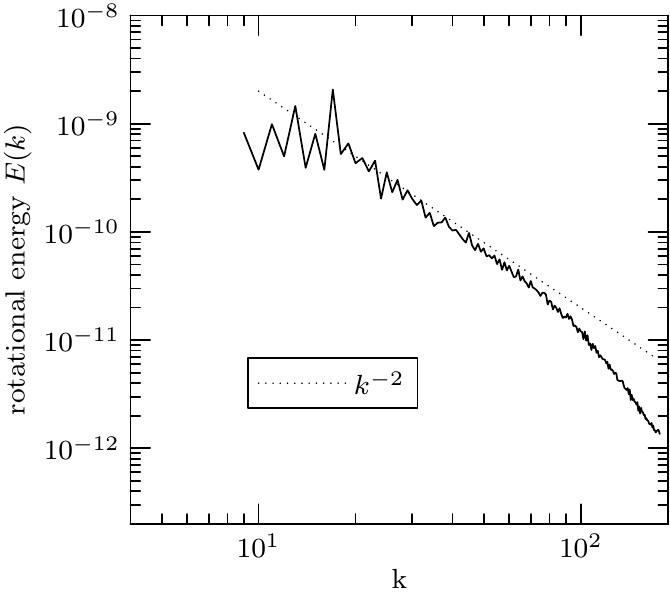}
\end{tabular}
\caption{\label{fig:energy} Turbulence with tolerance $\epsilon=5\times 10^{-2}$ for the inviscid and viscous runs.  Total energy minus the total energy at rest 
(left).  Energy spectrum for the rotational part of the velocity averaged over the interval $t=[132\mathrm{h}-136\mathrm{h}]$ (right).}
\end{figure}

The total energy $E(t)$ is defined as
\[
E(t) = \frac12 \int gh\left(gh + |u|^2 \right) \text{d}S - \frac12 c_{\text{wave}}^4 A_{S},
\]
where $A_S$ is the area of the sphere and the wave speed $c_{\text{wave}}$ is
\[
c_{\text{wave}} = \sqrt{\frac{g}{A_s}\int h \text{d}S}.
\]
Due to mass conservation, $c_{\text{wave}}$ is constant. Figure~\ref{fig:energy}~(left) shows that the total energy for both the viscous and inviscid runs first decreases and then stays at about the same level once the turbulence has developed.

The energy spectrum of the turbulent flows can be estimated by interpolating the adaptive results on a uniform grid and using spherical harmonics 
\[ f = \sum_{l=0}^N \sum_{m=-l}^l F_{lm} Y_{lm}. \]
The power spectrum is then defined as 
\[ S_f(l) = \sum_{m=-l}^l |F_{lm}|^2. \]
Figure~\ref{fig:energy} (right) show the spectrum of the rotational part of the velocity $\omega_v = \text{curl}_v~u$. The energy spectrum has a clearly defined power-law range, with a slope of about $-2.2$.

\section{Summary, conclusions and perspectives}	
This paper introduces a dynamically adaptive wavelet model for rotating shallow water equations on the sphere.  This model, based on the TRiSK discretization~\cite{Ringler10}, is an extension to spherical geometry of the method developed for the regular C-grid on the plane by \cite{DubKev13}.  The extension to the sphere is based on subdivisions of the icosahedron needs to overcome several challenges to cope with the irregular local C-grid geometry.  In addition to the extension to the sphere, the code has also been parallelized using {\tt mpi} using a highly efficient hybrid patch-tree data structure.  The {\tt metis}~\cite{metis} graph partitioner is used to improve load balancing amongst the cores. The model has been implemented in {\tt fortran95} in order to optimize computational efficiency.

The current implementation shows good strong parallel efficiency scaling for real test-cases up to $O(10^2)$ cores and good weak parallel efficiency scaling for load-balanced scenarios for up to at least $O(10^3)$ cores.  Acceptable parallel scaling to larger number of cores should be possible if the parallel  implementation is further optimized, for example by using measurement based multi-constraint load balancing or a hybrid shared/distributed memory approach.  Serial computational performance tests showed that the adaptive wavelet code is about 3 times slower than a non-adaptive TRiSK code and 5 times slower than a spectral solver per {\em active\/} node.  This suggests that the adaptive wavelet code should be faster than non-adaptive codes provided it achieves a grid compression ratio greater than 5.  However, the adaptive wavelet code also guarantees spatially uniform error control, which is not possible using non-adaptive methods.

The convergence, accuracy, error-control and efficiency properties of the adaptive wavelet method were confirmed using standard 
smooth test cases from \cite{Williamson} and a nonlinear unstable zonal jet test case proposed by~\cite{Galewsky04}. The method was 
also used to simulate viscous and inviscid fully-developed and decaying homogeneous and isotropic shallow water turbulence. Even in 
the challenging case of homogeneous turbulence the adaptive method was able to achieve high compression ratios of up to 15 to 50 
times due to the fine scale vorticity filaments that characterize the flow.  In this case, the wavelet method is 3 to 10 times faster than a 
spectral code with the same number of degrees of freedom. This suggests that the method should be appropriate for high Reynolds 
number geophysical flows without obvious large-scale sparsity.

To the best of our knowledge, the models in~\cite{StCyr2008Comparison}
are the only dynamically adaptive methods for the shallow water equations on the sphere comparable to the one we present here.  They analyze an interpolation-based spectral element shallow-water model on a cubed-sphere grid and a block-structured finite-volume method in latitude--longitude geometry.  It is instructive to compare and contrast our wavelet approach with these methods. 

In our case, the differential operators are discretized on an icosahedral grid using the TRiSK approximation proposed by \cite{Ringler10} to conserve important mimetic properties of the shallow water equations. The grid refinement guarantees a spatially uniform point-wise error estimated using wavelet coefficients, while \cite{StCyr2008Comparison} use an empirical refinement criterion.  When applied to the~\cite{Galewsky04} unstable zonal jet test problem our method requires roughly four to five times the number of degrees of freedom in order to obtain a similar quality of solution. This is likely due to the fact that the TRiSK scheme uses only second-order accurate approximations of the differential operators, while~\cite{StCyr2008Comparison} use fourth-order accurate approximations (at the cost of more computations per degree of freedom).  

\cite{StCyr2008Comparison} measure execution time for the adaptive mesh refinement (AMR) finite-volume code with three dyadic refinement levels  on 8, 16 and 24 cores.  They find that the AMR code is between 3.9 and 2.2 times slower than the fixed resolution code, similar to our overhead result with five refinement levels.  However, their strong parallel scaling appears to be weaker than in our case. The AMR code is only about 67\% efficient when increasing the number of cores from 8 to 24. In comparison, the adaptive wavelet code is over 95\% efficient for the same range of cores, and is 60\% efficient when comparing execution time on one core to execution time on 640 cores. 

Work is currently underway to incorporate coastlines and variable bathymetry with the short-term goal of developing a shallow water global oceans model.  This model will be applied both to tsunami propagation, and to the development and long-term dynamics of ocean flow, such as wind-driven gyres and western boundary currents.  In the medium-term, the model will be extended hydrostatically in the vertical direction while maintaining adaptivity based on horizontal structure.  The long-term goal of this work is to evaluate the potential of dynamically adaptive wavelet-based multiscale methods as dynamical cores for the next generation of climate and weather global circulation models.

\paragraph{Acknowledgements} NKRK would like to acknowledge to support of an NSERC Discovery grant, a Mobility Grant from the French Embassy in Ottawa and a visiting professorship at \'Ecole Polytechnique.  MA  acknowledges the support of a Mobility Grant from \'Ecole Polytechnique.

\bibliography{references}
\bibliographystyle{plain}
\end{document}